\documentclass[a4paper,12pt]{article}

\usepackage{amsmath,amssymb,amsfonts,amsthm} 
\usepackage{graphicx}
\usepackage{tikz}
\usepackage{color}
\usepackage{authblk}
\usepackage{verbatim}

\newtheorem{theorem}{Theorem}

\newcommand{\C}{\mathbb{C}}
\newcommand{\R}{\mathbb{R}}

\newcommand{\Z}{\mathbb{Z}}

\newcommand{\T}{\mathbb{T}}

\newcommand{\dd}{\mathrm{d}}

\newcommand{\ed}{\mathrm{e}}

\newcommand{\Const}{\mathrm{C}}
\newcommand{\const}{c}

\usepackage[margin=1cm,font=small]{caption}

\usepackage{layout} 

\newcommand{\coupling}{{\mathsf J}}  
\newcommand{\NumLevels}{{\mathsf N}}

\definecolor{col1}{rgb}{0,0,1}
\definecolor{col2}{rgb}{1,0,0.5}
\definecolor{col3}{rgb}{1,0.5,0.5}
\definecolor{col4}{rgb}{0.9,0.7,0.2}

\begin{document}

\title{Can translation invariant systems exhibit \\ a Many-Body Localized phase?}

\author[1]{Wojciech De Roeck} 
\affil[1]{\footnotesize{Instituut voor Theoretische Fysica, KU Leuven, Belgium,}

 \texttt{Wojciech.DeRoeck@fys.kuleuven.be}}

\author[2]{Fran\c cois Huveneers}
\affil[2]{\footnotesize{CEREMADE, Universit\' e Paris-Dauphine, France,}

\texttt{huveneers@ceremade.dauphine.fr}}

\date{\today}

\maketitle

\begin{abstract}
This note is based on a talk by one of us, F. H., at the conference PSPDE II, Minho 2013. We review some of our recent works related to (the possibility of) Many-Body Localization in the absence of quenched disorder, in particular \cite{1,2,3}. In these works, we provide arguments why systems without quenched disorder can exhibit `asymptotic' localization, but not genuine localization.  
\end{abstract}

\section{Introduction}

Anderson localization refers usually to the single-particle wave function of an electron being confined in some limited region of space for all times \cite{Anderson}. 
In perturbative regimes, being for example that of strong disorder, this phenomenon is by now well understood at both physical and mathematical levels \cite{Froehlich Spencer}.
The topic benefited recently from a huge revival of interest, when it was argued that localization could persist in presence of electron-electron interactions \cite{Basko Aleiner Altshuler}.   
Since in an interacting system, there is no appropriate analogue for one-particle wave functions, the phenomenon of Many-Body Localization (MBL) has to be characterized in other ways, than spatial decay of wave functions. 
One of the most appealing ways is to contrast MBL with an ergodic phase, where the latter can be defined as a phase in which the system thermalizes if initially prepared out of equilibrium.  
Several indicators can serve to quantify the ``absence of thermalization" and characterize the MBL phase:
\begin{enumerate}
\item
the existence of a complete set of integrals of motion, 
\item
an area law for the entanglement entropy of eigenstates,
\item  
the breakdown of the Eigenstate Thermalization Hypothesis (ETH),
\item 
the vanishing of transport coefficients (e.g.\@ thermal conductivity), etc...
\end{enumerate}
These characterizations are not equivalent (e.g.\@ while the first one fully describes the structure of the many-body Hilbert space, the last one implies only that relaxation times are sub-diffusive, strictly speaking); we refer to the recent review \cite{Nandkishore Huse review} for more explanations. 

Both one and many body localization can be understood as the result of frequency mismatch, or lack of resonances, between the different states of the system, in a similar way as the persistence of KAM tori can be understood in classical mechanics.
From a mathematical point of view, the study of MBL can be regarded as the investigation of the fate of KAM-like phenomena in the thermodynamic limit, where the volume is sent to infinity while the temperature is kept constant and positive. 
Recently, a mathematical approach based on the KAM scheme was indeed proposed to show the existence of MBL in one-dimensional spin chains \cite{Imbrie}.

\paragraph{MBL without disorder.}
As Anderson localization is naturally associated to an inhomogeneous medium, MBL has mostly been investigated for quenched disordered systems. 
Nevertheless, in a many-body set-up, interaction between particles could supply the needed frequency mismatch, 
as it is the case for the finite dimensional systems in the regime where KAM theory applies. 
The possibility of finding an MBL phase in translation invariant systems was recently explored by several authors
\cite{Kagan Maksimov}\cite{Grover Fisher}\cite{2}\cite{Schiulaz Mueller}\cite{Huse Nandkishore private}\cite{Hickey}.

To illustrate this, let us consider two classical models of coupled oscillator chains.
First, the one-dimensional pinned disordered harmonic chain, with Hamiltonian 
\begin{equation}\label{dhc}
H(q,p) \; = \; \frac{1}{2}\sum_{x} \big\{ p_x^2 + \omega_x^2 q_x^2 + \coupling (q_{x+1} - q_x)^2 \big\}, 
\qquad \omega_x \text{ i.i.d.}, \qquad \coupling > 0,
\end{equation}  
is a well known example of perfect thermal insulator, as a consequence of Anderson localization \cite{Bernardin Huveneers}.
Because the chain is harmonic, this system can indeed be seen as an example of one-particle localization rather than MBL.\footnote{
In fact, one believes that only quantum systems, e.g.\ like in \cite{Imbrie}, exhibit MBL in genuinely interacting systems.}
This, however, is as such not important for the point we want to make here:
to compare this system with the rotor chain, described by the Hamiltonian 
(with the angles $q_x \in \T = \R/(2\pi\Z)$ and the angular momenta $\omega_x\in\R$ canonical conjugates),
\begin{equation}\label{rotor}
H(\omega, q) \; = \; \frac{1}{2}\sum_{x} \big\{ \omega_x^2 + \coupling \cos (q_{x+1} - q_x) \big\} .
\end{equation}
In contrast to the disordered harmonic chain, $\omega_x$ are now dynamical variables rather than frozen degrees of freedom. 
Nevertheless, they can be regarded as i.i.d.\@ random variables with respect to the Gibbs state at positive temperature. 
Moreover, stability results for Hamiltonian mechanics at finite volume suggest that $\omega_x$ could be stable for very long or even infinite times.
Therefore, one might try to compare the high-temperature regime to the large disorder regime for disordered systems, 
and be led to the somehow provocative conjecture that thermal fluctuations can generate MBL.

\paragraph{Mobile bubbles.} 
Despite this suggestive analogy, the replacement of a fixed external disordered field by dynamical variables can in principle deeply modify the long time behavior of the system. 
Following \cite{3}\cite{Huse Nandkishore private}, it is the first purpose of this article to explore a mechanism for delocalization by means of rare mobile thermal bubbles 
that do not appear in quenched disorderd, fully MBL systems (having all their eigenstates localized), 
but that seem to be unavoidable in translation invariant systems with short range interactions.

We hope that these findings can also shed some light on the localization-delocalization transition in quenched disordered systems. 
First, in the cases where a transition in function of the energy density is expected \cite{Basko Aleiner Altshuler}, 
the argument developed here suggests in fact that there is no MBL phase at all (except possibly at zero temperature) in the true thermodynamic limit. 
Second, the absence of an MBL phase in classical disordered anharmonic systems could be explained by a mechanism of rare mobile ergodic bubbles too \cite{Oganesyan Pal Huse}\cite{Basko}.

At the moment of writing, the validity of our theoretical scenario is being investigated more thoroughly and also tested numerically \cite{De Roeck Huveneers Mueller Schiulaz}. 
Indeed, what our argument ultimately shows is that an overwhelming majority of the states in the Hilbert space are connected via a sequence of  resonant transitions (see below). 
Strictly speaking, delocalization does not follow at once; we refer to \cite{3} for a more detailed discussion on this point.

\paragraph{Asymptotic localization.} 
However, though true MBL can fail for the systems considered in this work, 
some asymptotic localization effects are surely expected, comparable to Nekhoroshev estimates at finite volume. 
We review here several recent mathematical works, 
where such estimates have in some sense been extended in the thermodynamic limit, 
predicting a non-analytic behavior of the thermal conductivity near the (trivial) critical point for various systems, 
as classical anharmonic disordered chains  \cite{Huveneers}, 
the rotor and DNLS chain \cite{1}, 
or a quantum particle system analogous to the Bose-Hubbard model \cite{2}.

\paragraph{Organization of the paper.}
The main quantum model studied in this paper is introduced in Section \ref{section: quenched and thermal}, where the principal difference between quenched and thermal disorder is discussed.
In Section \ref{rare bubbles perturbative}, a first version of the mechanism for delocalization, based on perturbation theory, is presented.
In Section \ref{section: bubbles non perturbative}, it is shown that non-perturbative considerations lead to the same conclusion, 
through a slightly less explicit but probably more robust way. 
Results on asymptotic localization are reviewed in Section \ref{section: asymptotic localization}.
Some technical issues are gathered in the appendices.

\section{Quenched and thermal disorder}\label{section: quenched and thermal}

We study quantum lattice systems defined in a large volume $V \subset \Z^d$ in the thermodynamic limit $|V|\to \infty$.
We think of each lattice site $x$ as containing a variable number of particles $\eta_x\in \{ 0, \dots , \NumLevels\}$, 
where $\NumLevels \ge 1$ is the maximal occupation number per site. 
There is thus a preferred product basis in the many-body Hilbert space, consisting of classical configurations  $|\eta \rangle = |(\eta_x)_{x}\rangle$. 
The total Hamiltonian is the sum of local operators: 
\begin{equation}\label{general sum of local terms Hamiltonian}
H = \sum_{A \subset \Z^d, |A| \le R} H_A \qquad \text{for some} \qquad R\ge 1,
\end{equation}
where $A$ are connected
and $H_A$ acts inside $A$, i.e.,
\begin{equation}\label{set dependence H A}
\langle \eta' | H_A | \eta \rangle \; = \; 0 
\qquad \text{if} \qquad 
\eta_x' \ne \eta_x \quad \text{for} \quad x \notin A. 
\end{equation} 
We assume that the total number of particles is conserved: $[H,N]=0$ with $N=\sum_{x}b_x^* b_x$, 
where $b_x$ and $b_x^*$ are respectively bosonic annihilation and creation operators with cut-off $\NumLevels$: 
$b_x | \dots , \eta_x, \dots \rangle = \sqrt{\eta_x} | \dots , \eta_{x} - 1, \dots \rangle$ if $1 \le \eta_x \le \NumLevels$ 
and $b_x | \eta \rangle = 0$ if $\eta_x = 0$.
Moreover, we consider the case where $H$ is a small perturbation of a Hamiltonian $H^{(0)}$ which is manifestly MBL, 
for example $H^{(0)}$ of the type \eqref{general sum of local terms Hamiltonian} and diagonal in the $\{ | \eta \rangle \}$ basis, 
\begin{equation}\label{general perturbative Hamiltonian}
H \; = \; H^{(0)} + \coupling H^{(1)}, \qquad \coupling \; \ll \; 1.
\end{equation}
More precisely, we require that the matrix elements of $\coupling H^{(1)}$ are smaller than the level spacing of the local Hamiltonians $H_A$.

\begin{figure}[h!]
\begin{center}
\begin{tikzpicture}[scale=0.7]

\begin{scope}[xshift=-4.2cm]
\draw [>=stealth,->] (-1.2,0) -- (6.2,0);
\draw [>=stealth,->] (-1,-0.2) -- (-1,3.2);

\draw (5.7,-0.1) node[below]{$x$} ;
\draw (-1.1,2.9) node[left]{$\eta_x$} ;

\draw [ultra thick] (-0.9,1) -- (-0.1,1);
\draw [ultra thick] (0.1,2) -- (0.9,2);
\draw [ultra thick] (3.1,1.5) -- (3.9,1.5);
\draw [ultra thick] (4.1,2.5) -- (4.9,2.5);
\draw [ultra thick] (5.1,1) -- (5.9,1);

\draw [ultra thick,dashed] (-0.9,0.5) -- (-0.1,0.5);
\draw [ultra thick,dashed] (0.1,2.5) -- (0.9,2.5);
\draw [ultra thick,dashed] (3.1,1) -- (3.9,1);
\draw [ultra thick,dashed] (5.1,1.5) -- (5.9,1.5);

\draw [>=stealth,->] (-0.5,1) -- (-0.5,0.6);
\draw [>=stealth,->] (0.5,2) -- (0.5,2.4);
\draw [>=stealth,->] (3.5,1.5) -- (3.5,1.1);
\draw [>=stealth,->] (5.5,1) -- (5.5,1.4);

\draw [ultra thick] (-1.1,0.5) -- (-1,0.5);
\draw [ultra thick] (-1.1,1) -- (-1,1);
\draw [ultra thick] (-1.1,1.5) -- (-1,1.5);
\draw [ultra thick] (-1.1,2) -- (-1,2);
\draw [ultra thick] (-1.1,2.5) -- (-1,2.5);

\draw[thick,dotted] (1.77,1.5) -- (2.23,1.5);
\end{scope}

\begin{scope}[xshift=4.2cm]
\draw [>=stealth,->] (-0.2,0) -- (6,0);
\draw [>=stealth,->] (0,-0.2) -- (0,3.3);

\draw (5.7,-0.1) node[below]{$x$} ;
\draw (-0.1,0.6) node[left]{$|\eta'\rangle$} ;
\draw (-0.1,2.7) node[left]{$|\eta\rangle$} ;

\draw [thick,dotted] (0.1,2.625) -- (0.45,2.625); 
\draw [ultra thick]  (0.55, 2.5) -- (0.95,2.5);
\draw [ultra thick,blue]  (1.05, 3) -- (1.45,3);
\draw [ultra thick,blue]  (1.55, 2.75) -- (1.95,2.75);
\draw [ultra thick,blue]  (2.05, 2.75) -- (2.45,2.75);
\draw [ultra thick,blue]  (2.55, 2.5) -- (2.95,2.5);
\draw [ultra thick,blue]  (3.05, 2.25) -- (3.45,2.25);
\draw [ultra thick]  (3.55, 3) -- (3.95,3);
\draw [ultra thick]  (4.05, 2.5) -- (4.45,2.5);
\draw [ultra thick]  (4.55, 2.75) -- (4.95,2.75);
\draw [ultra thick]  (5.05, 2.25) -- (5.45,2.25);
\draw [thick,dotted]  (5.55, 2.625) -- (5.9,2.625);

\draw [thick,dotted] (0.1,0.625) -- (0.45,0.625); 
\draw [ultra thick,blue]  (0.55, 0.5) -- (0.95,0.5);
\draw [ultra thick,blue]  (1.05, 0.25) -- (1.45,0.25);
\draw [ultra thick,blue]  (1.55, 0.5) -- (1.95,0.5);
\draw [ultra thick,blue]  (2.05, 0.75) -- (2.45,0.75);
\draw [ultra thick,blue]  (2.55, 0.75) -- (2.95,0.75);
\draw [ultra thick,blue]  (3.05, 1) -- (3.45,1);
\draw [ultra thick,blue]  (3.55, 1) -- (3.95,1);
\draw [ultra thick]  (4.05, 0.5) -- (4.45,0.5);
\draw [ultra thick]  (4.55, 0.75) -- (4.95,0.75);
\draw [ultra thick]  (5.05, 0.25) -- (5.45,0.25);
\draw [thick,dotted]  (5.55, 0.625) -- (5.9,0.625);
\end{scope}
\end{tikzpicture}
\end{center}

\caption{
\label{figure: resonances tranlsation invariant Hamiltonian}
First order resonances for the Hamiltonian \eqref{Bose Hubbard Hamiltonian}. 
Left panel. 
Non-resonant nearest-neighbor hopping on the left; resonant second neighbor hopping on the right. 
Right panel. 
As shown in Appendix 1, the classical state $|\eta\rangle$ is connected to $|\eta'\rangle$ through a sequence of resonant transitions. 
The naive resonant spot in $|\eta\rangle$, indicated in blue, appears thus as part of a larger one seen in $|\eta'\rangle$. 
}
\end{figure}
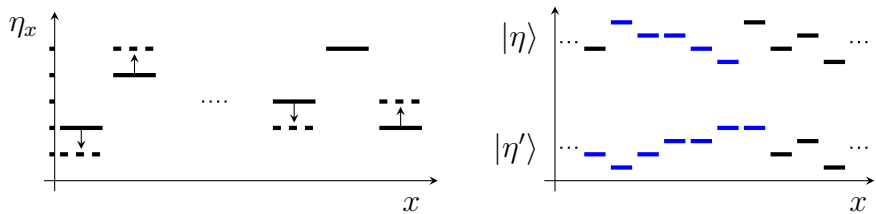

Following \cite{3}, we give two examples. 
In the first example, translation invariance is broken by an external field (quenched disorder). 
At small enough coupling $\coupling >0$, 
resonances, that are potentially responsible for delocalization in the considered perturbative regime, are rare and their location is determined by the external field. 
In the second example, translation invariance is restored.
Although,  for small values of $\coupling$, resonances are equally rare in the second example as in the first one, 
they can appear everywhere and therefore they can be possibly part of a giant cluster in configuration space 
(see the right panel in figure \ref{figure: resonances tranlsation invariant Hamiltonian}). 
This is the main difference between these two types of systems.

\paragraph{Quenched disordered Hamiltonian.}
We let $d=1$, $\NumLevels = 1$ (i.e.\ we have hard-core bosons) and we define a Hamiltonian of the form \eqref{general perturbative Hamiltonian} by
\begin{equation}\label{Hamiltonian disordered chain}
H 
\; = \;
 \sum_x \omega_x b_x^* b_x \, + \, \coupling \sum_x \big\{ b_x^* (b_{x+1} + b_{x+2}) + b_x  (b_{x+1}^* + b_{x+2}^*) \big\}, 
\qquad \omega_x \text{ i.i.d.}
\end{equation}
(second neighbor hopping is introduced to break integrability). 
We simply refer to $H^{(0)}$ as the uncoupled energy, and to $H^{(1)}$ as the hopping term. 
To facilitate the comparison with the translation invariant case described below,
we assume that the values of $\omega_x$ are uniformly distributed in a finite set
$\{ \alpha_0, \dots , \alpha_{\NumLevels'} \}$ for some $\NumLevels' \gg 1$. 
We think of the numbers $\alpha_0 < \dots < \alpha_{\NumLevels'}$ as picked at random in an interval of size 1, so that $\alpha_{k+1} - \alpha_k \sim 1/\NumLevels'$.   
We assume $\coupling \ll 1/\NumLevels'$. 

Consider now two classical states $| \eta \rangle$ and $| \eta' \rangle$ connected in the first order in perturbation in $\coupling$: $\langle \eta' | H^{(1)} | \eta \rangle \ne 0$. 
These two states only differ from each other by one particle that hopped one or two sites away. 
Thanks to the properties of the random field $(\omega_x)_x$ described above, at most of the places, 
the change in energy due to the hopping cannot compensate the resulting change in uncoupled energy: 
\begin{equation}\label{non resonance inequality}
\coupling \big|\langle \eta' | H^{(1)} | \eta \rangle\big| \; \ll \; \big| H^{(0)} (\eta') -  H^{(0)} (\eta)  \big|
\, = \, |\omega_y - \omega_x|, 
\end{equation}
assuming that the particle hops from $x$ to $y$, with $|y-x|\le 2$
(here and below, the convention $A (\phi) = \langle \phi | A | \phi \rangle$ is used whenever $\phi$ is an eigenstate of $A$).
As a consequence, it is possible to remove these transitions via a perturbative procedure \cite{Imbrie}.
However, in some rare places, the hopping does not produce any change in the uncoupled energy: 
\begin{equation}\label{resonance disordered case}
H^{(0)} (\eta')  -  H^{(0)} (\eta)  \; = \;  \omega_y - \omega_x  \; = \; 0. 
\end{equation}
In these cases, despite the fact that $\coupling \ll 1$, the hopping becomes the dominant effect and cannot longer be treated perturbatively.
These latter transitions are called resonant. 
However, the most important point seen from \eqref{resonance disordered case}, is that the location of resonances is determined by the external field $(\omega_x)_x$ alone:
the second equality in \eqref{resonance disordered case} only depends on the points $x$ and $y$, and not on the configurations $| \eta \rangle$ and $| \eta' \rangle$.    
Moreover, the probability with respect to the distribution of $(\omega_x)_x$, of finding a resonance somewhere is of the order of $1/\NumLevels'$, so that 
even for $d > 1$, they form small isolated clusters in physical space. 
This is the reason why they entail no delocalization.

\paragraph{Translation invariant Hamiltonian.}
We let $d=1$, $\NumLevels \gg 1$ and we consider the Bose-Hubbard Hamiltonian with the cut-off  $\NumLevels$ on the number of particles per site, and with second neighbor hopping: 
\begin{equation}\label{Bose Hubbard Hamiltonian}
H \; = \; \sum_{x} (b_x^* b_x)^2 \, + \,  \coupling \sum_x \big\{ b_x^* (b_{x+1} + b_{x+2}) + b_x  (b_{x+1}^* + b_{x+2}^*) \big\} . 
\end{equation}
This Hamiltonian is of the form \eqref{general perturbative Hamiltonian}:
$H^{(0)}$creates a repulsion between particles, and is referred to as the interaction energy, while $H^{(1)}$ allows again for particles to hop. 

First order resonances in $\coupling$ are defined as in the disordered case:
two classical states $| \eta \rangle$ and $| \eta' \rangle$ such that $\langle \eta' | H^{(1)} | \eta\rangle \ne 0$, 
enter in resonance if the inequality in \eqref{non resonance inequality} is violated.
Thanks to the strong anharmonicity in the interaction, coming from the square in $(b_x^* b_x)^2$, 
and thanks to the assumption $\coupling \ll 1$,
a resonance between the states $|\eta \rangle$ and $| \eta' \rangle$, due to the hopping of a particle from $x$ to $y$, with $|x-y| \le 2$, only occurs if 
\begin{equation}\label{resonance translation invariant}
\eta_x'    \; = \;  \eta_{y} \; = \; \eta_x + 1 \; = \;  \eta_{y}' +1 \qquad \text{or}  \qquad  \eta_x'    \; = \;  \eta_{y} \; = \; \eta_x - 1 \; = \;  \eta_{y}' -1.  
\end{equation}
This is illustrated on the left panel of figure \ref{figure: resonances tranlsation invariant Hamiltonian}.

For the disordered Hamiltonian considered above, 
the distribution of the disorder allowed to quantify how likely it was for a hopping between two given sites to induce a resonant transition.
This comes as a little surprise on second thought: the inequality in \eqref{non resonance inequality} involves two states $| \eta \rangle$ and $| \eta' \rangle$
so that we would naturally expect the nature of the transition (resonant or not) to depend on these states.
In general, the Gibbs state can serve to quantify the occurrence of resonances.
Here for example, thanks to \eqref{resonance translation invariant}, 
we can tell how likely it is for a resonance to occur due to a hop between $x$ and $y$, for a typical classical state $| \eta \rangle$ in the Gibbs state.
To simplify the discussion, we consider the Gibbs state at infinite temperature, 
such that, in the basis of classical configurations, it can be viewed as a bonafide probability measure that gives equal weight to each configuration, 
allowing to define unambiguously the probability of a resonance. 
However, as the number of particles is conserved, we can and will fix the density $\rho$ of particles per site.
By definition, $0 \le \rho \le \NumLevels$ is the average number of particles per site.

We distinguish two regimes. 
First, at very low densities\footnote{An analogous conclusion actually also holds at very large densities due to the cut-off $\NumLevels$.} ($\rho \ll 1$), 
most of the particles are isolated, so that, according to \eqref{resonance translation invariant}, they can typically hop via resonant transitions. 
An ergodic phase is expected, though, since $d=1$, atypical clusters of particles can slow down thermalization very much \cite{Carleo et al}.  
Second, at densities close to $\NumLevels/2$, 
we deduce from \eqref{resonance translation invariant} that the probability of finding a resonance somewhere becomes of the order of $1/\NumLevels$, 
precisely as in the quenched disordered case, with $\NumLevels'$ there being $\NumLevels$ here.
This observation is a possible starting point to address the question of Many-Body Localization \cite{2}\cite{Schiulaz Mueller}\cite{3}. 
However, due to translation invariance, resonant spots have no preferred location. 
Therefore, there is no reason why the picture of small isolated clusters would remain valid. 
Contrary to what a naive look at typical classical configurations suggests,
resonant spots might be just the ``visible part of the iceberg", a giant cluster in configuration space, 
as illustrated on the right panel of figure \ref{figure: resonances tranlsation invariant Hamiltonian}.

\section{Rare bubbles due to perturbative effects}\label{rare bubbles perturbative}

Following \cite{3}, 
we here study more systematically the effect illustrated on the right panel of figure \ref{figure: resonances tranlsation invariant Hamiltonian}, 
though we switch to $d=2$ as that allows for a more transparent discussion. 
More precisely, we show that, in the thermodynamic limit, the overwhelming majority of pairs of classical configurations are connected 
via a sequence of resonant transitions in the first order in $\coupling$, 
as long as some obvious conservation laws are satisfied (resonant transitions preserve the number of sites with a given occupation number). 
Therefore, we conclude that the choice of a typical classical state as initial state, does not suffice for the picture of small resonant localized islands to survive. 

To achieve this, we observe that, in a typical classical state at some density $\rho \sim \NumLevels/2$, arbitrary large low density bubbles are found in the thermodynamic limit.
Since, as noticed in the previous section, low density states are expected to be ergodic for the Hamiltonian \eqref{Bose Hubbard d=2} below, 
it should be possible to displace these bubbles across the system through a sequence of resonant transitions in the first order in $\coupling$, 
at least if this phenomenon is already visible in a first order approximation. 
We prove below that this is indeed the case, by showing that resonant transitions allow the bubble to absorb and expel particles from its surrounding. 
By continuously absorbing and expelling, it can thus move across the system and carry particles all over the places, 
providing a mechanism to connect all states satisfying the above mentioned conservation laws.\footnote
{  
One could wonder whether localization could emerge at some larger scale. 
As all involved transitions are resonant, there is surely no obvious reason to think that bubbles are localized. 
Nevertheless, we will see that a long specific sequence of transitions is required for a bubble to move over a distance of order one, 
as, for a random sequence of the same length, the bubble would rather shrink to a minimal size than start moving.
This feature could suggest that the bubbles are ultimately localized, though no conclusion in that direction can be drawn from this observation alone.
}

\paragraph{The resonant Hamiltonian.}
Let $d=2$ and $\NumLevels \gg 1$. 
We consider the Bose-Hubbard Hamiltonian with the cut-off  $\NumLevels$ on the number of particles per site:
\begin{equation}\label{Bose Hubbard d=2}
H \; = \; 
H^{(0)} + \coupling H^{(1)}
\; = \; 
\sum_x (b_x^* b_x )^2 \, + \, \frac{\coupling}{2} \sum_{x,y:x \sim y} \big\{ b_x^* b_y + b_x b_y^* \big\}, 
\end{equation}
where $x\sim y$ means $|x-y|_1 = 1$ (nearest neighbors).
Transitions that are non-resonant in the first order in $\coupling$ can be removed perturbatively, up to second order corrections \cite{Imbrie}. 
Therefore, if these corrections are neglected in a first approximation, the dynamics is dominated by the resonant Hamiltonian $H_{res}$ defined by
\begin{align}
\langle \eta' | H_{res} | \eta \rangle 
\; &=\;
\langle \eta' | H^{(0)} | \eta \rangle 
\, + \, 
 \frac{\coupling}{2}  \delta \left( H^{(0)}(\eta') - H^{(0)}(\eta) \right) \langle \eta' | H^{(1)} | \eta \rangle
\nonumber\\
\; &= \;
\langle \eta' | H^{(0)} | \eta \rangle \,+ \, \coupling \sum_{x,y:x \sim y}  \delta (\eta_y - \eta_x - 1) \langle \eta' | b_x^* b_y | \eta \rangle, 
\label{resonant Bose Hubbard d=2}
\end{align} 
with $\delta (\cdot)$ the Kronecker delta function. 
Let us make some remarks:
\begin{enumerate}
\item
The fact that two classical states $| \eta\rangle$ and $| \eta' \rangle$ are connected via a sequence of resonant transitions, 
can now  simply be expressed by saying that $|\eta \rangle$ and $| \eta' \rangle$ are connected via $H_{res}$, meaning  
\begin{equation}\label{connected via Hamiltonian}
\langle \eta' | \ed^{-it H_{res}} | \eta \rangle \; \ne \; 0 \qquad \text{for some} \qquad t\ge 0.
\end{equation}

\item
By definition of $H_{res}$, if two states are connected via $H_{res}$, then they have the same value of $H^{(0)}$. 
Therefore, the term $\langle \eta' | H^{(0)} | \eta \rangle$ in \eqref{resonant Bose Hubbard d=2} is constant in every subspace left invariant by $H_{res}$, 
and could thus be removed. This confirms that resonant transitions are non-perturbative even though $\coupling \ll 1$. 

\item
If the dynamics generated by $H_{res}$ is delocalized, it is hard to conceive that higher orders could restore localization. 
On the other hand, 
should the dynamics generated by $H_{res}$ be localized, we would need to check the effect of higher orders before we conclude that there is localization.
\end{enumerate}

\paragraph{The bubble.}
Let us consider a classical initial state $|\eta\rangle$ that is typical at the density $\rho \sim \NumLevels/2$ in the thermodynamic limit,
and let $\mathcal B \subset V$ be a connected region corresponding to a bubble with density $\tilde\rho$ so small that $\tilde\rho \ll 1$. 
Since $\NumLevels \gg 1$, the number of particles sitting on a given site in $\mathcal B$, is approximately given by an exponential distribution with mean $\tilde\rho$.
We assume that $|\mathcal B|$ is large enough so that it is likely to find at least one site occupied by $\NumLevels$ particles in $\mathcal B$
($|\mathcal B| \gtrsim \ed^{\NumLevels/{\tilde\rho}}$ is thus needed). 
Let us denote by $n_k$ be the number of sites in $\mathcal B$ occupied by $k$ particles ($0 \le k \le \NumLevels$).
There exists a number $p\sim \ed^{1/\tilde{\rho}} \gg 1$ so that $n_k \le (1/p) n_{k-1}$ for $1 \le k \le \NumLevels$. 

For convenience, let us assume that $\mathcal B$ has a rectangular shape, as depicted on figure \ref{figure: flemish mountain}.
Let us first show that, inside the bubble, it is possible to move the particles all over the places via a sequence of resonant transitions. 
To go directly to the point, let us consider an initial classical state $| \eta \rangle$ such as depicted on figure \ref{figure: flemish mountain}, 
with high density sites in the upper left corner (our argument shows that we could well have started from a more typical configuration inside the bubble).  
On figure \ref{figure: flemish mountain}, we show a systematic strategy to connect this state via a sequence of resonant transitions (see \eqref{connected via Hamiltonian}) to another classical state $|\eta'\rangle$ where the high density sites are in the lower right corner. 
This is exemplary: it is seen that the sites with $\NumLevels$ particles (the blue square on figure \ref{figure: flemish mountain}) could first be placed everywhere, 
and that sites with $\NumLevels - 1$ particles (the dark pink squares) could then in turn be placed at (almost) any of the remaining places, etc...\footnote
{It is well possible that, due to geometrical constraints, not precisely all configurations can be reached that way, but this is not relevant for the discussion.}

As a consequence, starting from a classical configuration $| \eta \rangle$ inside the bubble, it is possible to populate any given site on the inside border of the bubble by any given number of particles, via resonant transitions only. 
Therefore, it is also possible to absorb a particle on a site on the outside border inside the bubble via a sequence of resonant transitions, or to expel a particle outside.   
We conclude that the bubble can move accross the system, and carry particles from any site to any site, via a sequence of resonant transitions.

\begin{figure}[h!]
\begin{center}
\begin{tikzpicture}[scale=1]

\draw (0,0) node{{\color{col4}$\blacksquare$}};
\draw (0.3,0) node{{\color{col4}$\blacksquare$}};
\draw (0.6,0) node{{\color{col4}$\blacksquare$}};
\draw (0.9,0) node{{\color{col4}$\blacksquare$}};
\draw (1.2,0) node{{\color{col4}$\blacksquare$}};
\draw (1.5,0) node{{\color{col4}$\blacksquare$}};
\draw (1.8,0) node{{\color{col4}$\blacksquare$}};
\draw (2.1,0) node{{\color{col4}$\blacksquare$}};
\draw (2.4,0) node{{\color{col4}$\blacksquare$}};
\draw (2.7,0) node{{\color{col4}$\blacksquare$}};
\draw (3,0) node{{\color{col4}$\blacksquare$}};
\draw (3.3,0) node{{\color{col4}$\blacksquare$}};
\draw (3.6,0) node{{\color{col4}$\blacksquare$}};
\draw (3.9,0) node{{\color{col4}$\blacksquare$}};
\draw (4.2,0) node{{\color{col4}$\blacksquare$}};

\draw (0,0.3) node{{\color{col3}$\blacksquare$}};
\draw (0.3,0.3) node{{\color{col3}$\blacksquare$}};
\draw (0.6,0.3) node{{\color{col3}$\blacksquare$}};
\draw (0.9,0.3) node{{\color{col3}$\blacksquare$}};
\draw (1.2,0.3) node{{\color{col3}$\blacksquare$}};
\draw (1.5,0.3) node{{\color{col3}$\blacksquare$}};
\draw (1.8,0.3) node{{\color{col3}$\blacksquare$}};
\draw (2.1,0.3) node{{\color{col4}$\blacksquare$}};
\draw (2.4,0.3) node{{\color{col4}$\blacksquare$}};
\draw (2.7,0.3) node{{\color{col4}$\blacksquare$}};
\draw (3,0.3) node{{\color{col4}$\blacksquare$}};
\draw (3.3,0.3) node{{\color{col4}$\blacksquare$}};
\draw (3.6,0.3) node{{\color{col4}$\blacksquare$}};
\draw (3.9,0.3) node{{\color{col4}$\blacksquare$}};
\draw (4.2,0.3) node{{\color{col4}$\blacksquare$}};

\draw (0,0.6) node{{\color{col2}$\blacksquare$}};
\draw (0.3,0.6) node{{\color{col2}$\blacksquare$}};
\draw (0.6,0.6) node{{\color{col2}$\blacksquare$}};
\draw (0.9,0.6) node{{\color{col3}$\blacksquare$}};
\draw (1.2,0.6) node{{\color{col3}$\blacksquare$}};
\draw (1.5,0.6) node{{\color{col3}$\blacksquare$}};
\draw (1.8,0.6) node{{\color{col3}$\blacksquare$}};
\draw (2.1,0.6) node{{\color{col4}$\blacksquare$}};
\draw (2.4,0.6) node{{\color{col4}$\blacksquare$}};
\draw (2.7,0.6) node{{\color{col4}$\blacksquare$}};
\draw (3,0.6) node{{\color{col4}$\blacksquare$}};
\draw (3.3,0.6) node{{\color{col4}$\blacksquare$}};
\draw (3.6,0.6) node{{\color{col4}$\blacksquare$}};
\draw (3.9,0.6) node{{\color{col4}$\blacksquare$}};
\draw (4.2,0.6) node{{\color{col4}$\blacksquare$}};

\draw (0,0.9) node{{\color{col1}$\blacksquare$}};
\draw (0.3,0.9) node{{\color{col2}$\blacksquare$}};
\draw (0.6,0.9) node{{\color{col2}$\blacksquare$}};
\draw (0.9,0.9) node{{\color{col3}$\blacksquare$}};
\draw (1.2,0.9) node{{\color{col3}$\blacksquare$}};
\draw (1.5,0.9) node{{\color{col3}$\blacksquare$}};
\draw (1.8,0.9) node{{\color{col3}$\blacksquare$}};
\draw (2.1,0.9) node{{\color{col4}$\blacksquare$}};
\draw (2.4,0.9) node{{\color{col4}$\blacksquare$}};
\draw (2.7,0.9) node{{\color{col4}$\blacksquare$}};
\draw (3,0.9) node{{\color{col4}$\blacksquare$}};
\draw (3.3,0.9) node{{\color{col4}$\blacksquare$}};
\draw (3.6,0.9) node{{\color{col4}$\blacksquare$}};
\draw (3.9,0.9) node{{\color{col4}$\blacksquare$}};
\draw (4.2,0.9) node{{\color{col4}$\blacksquare$}};

\begin{scope}[xshift=5.2cm]
\draw (0,0) node{{\color{col4}$\blacksquare$}};
\draw (0.3,0) node{{\color{col4}$\blacksquare$}};
\draw (0.6,0) node{{\color{col4}$\blacksquare$}};
\draw (0.9,0) node{{\color{col4}$\blacksquare$}};
\draw (1.2,0) node{{\color{col4}$\blacksquare$}};
\draw (1.5,0) node{{\color{col4}$\blacksquare$}};
\draw (1.8,0) node{{\color{col4}$\blacksquare$}};
\draw (2.1,0) node{{\color{col4}$\blacksquare$}};
\draw (2.4,0) node{{\color{col4}$\blacksquare$}};
\draw (2.7,0) node{{\color{col4}$\blacksquare$}};
\draw (3,0) node{{\color{col4}$\blacksquare$}};
\draw (3.3,0) node{{\color{col4}$\blacksquare$}};
\draw (3.6,0) node{{\color{col4}$\blacksquare$}};
\draw (3.9,0) node{{\color{col4}$\blacksquare$}};
\draw (4.2,0) node{{\color{col4}$\blacksquare$}};

\draw (0,0.3) node{{\color{col3}$\blacksquare$}};
\draw (0.3,0.3) node{{\color{col3}$\blacksquare$}};
\draw (0.6,0.3) node{{\color{col3}$\blacksquare$}};
\draw (0.9,0.3) node{{\color{col3}$\blacksquare$}};
\draw (1.2,0.3) node{{\color{col3}$\blacksquare$}};
\draw (1.5,0.3) node{{\color{col3}$\blacksquare$}};
\draw (1.8,0.3) node{{\color{col3}$\blacksquare$}};
\draw (2.1,0.3) node{{\color{col4}$\blacksquare$}};
\draw (2.4,0.3) node{{\color{col4}$\blacksquare$}};
\draw (2.7,0.3) node{{\color{col4}$\blacksquare$}};
\draw (3,0.3) node{{\color{col4}$\blacksquare$}};
\draw (3.3,0.3) node{{\color{col4}$\blacksquare$}};
\draw (3.6,0.3) node{{\color{col4}$\blacksquare$}};
\draw (3.9,0.3) node{{\color{col4}$\blacksquare$}};
\draw (4.2,0.3) node{{\color{col4}$\blacksquare$}};

\draw (0,0.6) node{{\color{col2}$\blacksquare$}};
\draw (0.3,0.6) node{{\color{col2}$\blacksquare$}};
\draw (0.6,0.6) node{{\color{col1}$\blacksquare$}};
\draw (0.9,0.6) node{{\color{col3}$\blacksquare$}};
\draw (1.2,0.6) node{{\color{col3}$\blacksquare$}};
\draw (1.5,0.6) node{{\color{col3}$\blacksquare$}};
\draw (1.8,0.6) node{{\color{col3}$\blacksquare$}};
\draw (2.1,0.6) node{{\color{col4}$\blacksquare$}};
\draw (2.4,0.6) node{{\color{col4}$\blacksquare$}};
\draw (2.7,0.6) node{{\color{col4}$\blacksquare$}};
\draw (3,0.6) node{{\color{col4}$\blacksquare$}};
\draw (3.3,0.6) node{{\color{col4}$\blacksquare$}};
\draw (3.6,0.6) node{{\color{col4}$\blacksquare$}};
\draw (3.9,0.6) node{{\color{col4}$\blacksquare$}};
\draw (4.2,0.6) node{{\color{col4}$\blacksquare$}};

\draw (0,0.9) node{{\color{col2}$\blacksquare$}};
\draw (0.3,0.9) node{{\color{col2}$\blacksquare$}};
\draw (0.6,0.9) node{{\color{col2}$\blacksquare$}};
\draw (0.9,0.9) node{{\color{col3}$\blacksquare$}};
\draw (1.2,0.9) node{{\color{col3}$\blacksquare$}};
\draw (1.5,0.9) node{{\color{col3}$\blacksquare$}};
\draw (1.8,0.9) node{{\color{col3}$\blacksquare$}};
\draw (2.1,0.9) node{{\color{col4}$\blacksquare$}};
\draw (2.4,0.9) node{{\color{col4}$\blacksquare$}};
\draw (2.7,0.9) node{{\color{col4}$\blacksquare$}};
\draw (3,0.9) node{{\color{col4}$\blacksquare$}};
\draw (3.3,0.9) node{{\color{col4}$\blacksquare$}};
\draw (3.6,0.9) node{{\color{col4}$\blacksquare$}};
\draw (3.9,0.9) node{{\color{col4}$\blacksquare$}};
\draw (4.2,0.9) node{{\color{col4}$\blacksquare$}};
\end{scope}

\begin{scope}[yshift=-1.7cm]
\draw (0,0) node{{\color{col4}$\blacksquare$}};
\draw (0.3,0) node{{\color{col4}$\blacksquare$}};
\draw (0.6,0) node{{\color{col4}$\blacksquare$}};
\draw (0.9,0) node{{\color{col4}$\blacksquare$}};
\draw (1.2,0) node{{\color{col4}$\blacksquare$}};
\draw (1.5,0) node{{\color{col4}$\blacksquare$}};
\draw (1.8,0) node{{\color{col4}$\blacksquare$}};
\draw (2.1,0) node{{\color{col4}$\blacksquare$}};
\draw (2.4,0) node{{\color{col4}$\blacksquare$}};
\draw (2.7,0) node{{\color{col4}$\blacksquare$}};
\draw (3,0) node{{\color{col4}$\blacksquare$}};
\draw (3.3,0) node{{\color{col4}$\blacksquare$}};
\draw (3.6,0) node{{\color{col4}$\blacksquare$}};
\draw (3.9,0) node{{\color{col4}$\blacksquare$}};
\draw (4.2,0) node{{\color{col4}$\blacksquare$}};

\draw (0,0.3) node{{\color{col3}$\blacksquare$}};
\draw (0.3,0.3) node{{\color{col3}$\blacksquare$}};
\draw (0.6,0.3) node{{\color{col2}$\blacksquare$}};
\draw (0.9,0.3) node{{\color{col2}$\blacksquare$}};
\draw (1.2,0.3) node{{\color{col2}$\blacksquare$}};
\draw (1.5,0.3) node{{\color{col3}$\blacksquare$}};
\draw (1.8,0.3) node{{\color{col3}$\blacksquare$}};
\draw (2.1,0.3) node{{\color{col4}$\blacksquare$}};
\draw (2.4,0.3) node{{\color{col4}$\blacksquare$}};
\draw (2.7,0.3) node{{\color{col4}$\blacksquare$}};
\draw (3,0.3) node{{\color{col4}$\blacksquare$}};
\draw (3.3,0.3) node{{\color{col4}$\blacksquare$}};
\draw (3.6,0.3) node{{\color{col4}$\blacksquare$}};
\draw (3.9,0.3) node{{\color{col4}$\blacksquare$}};
\draw (4.2,0.3) node{{\color{col4}$\blacksquare$}};

\draw (0,0.6) node{{\color{col3}$\blacksquare$}};
\draw (0.3,0.6) node{{\color{col3}$\blacksquare$}};
\draw (0.6,0.6) node{{\color{col1}$\blacksquare$}};
\draw (0.9,0.6) node{{\color{col2}$\blacksquare$}};
\draw (1.2,0.6) node{{\color{col2}$\blacksquare$}};
\draw (1.5,0.6) node{{\color{col3}$\blacksquare$}};
\draw (1.8,0.6) node{{\color{col3}$\blacksquare$}};
\draw (2.1,0.6) node{{\color{col4}$\blacksquare$}};
\draw (2.4,0.6) node{{\color{col4}$\blacksquare$}};
\draw (2.7,0.6) node{{\color{col4}$\blacksquare$}};
\draw (3,0.6) node{{\color{col4}$\blacksquare$}};
\draw (3.3,0.6) node{{\color{col4}$\blacksquare$}};
\draw (3.6,0.6) node{{\color{col4}$\blacksquare$}};
\draw (3.9,0.6) node{{\color{col4}$\blacksquare$}};
\draw (4.2,0.6) node{{\color{col4}$\blacksquare$}};

\draw (0,0.9) node{{\color{col3}$\blacksquare$}};
\draw (0.3,0.9) node{{\color{col3}$\blacksquare$}};
\draw (0.6,0.9) node{{\color{col3}$\blacksquare$}};
\draw (0.9,0.9) node{{\color{col3}$\blacksquare$}};
\draw (1.2,0.9) node{{\color{col3}$\blacksquare$}};
\draw (1.5,0.9) node{{\color{col3}$\blacksquare$}};
\draw (1.8,0.9) node{{\color{col3}$\blacksquare$}};
\draw (2.1,0.9) node{{\color{col4}$\blacksquare$}};
\draw (2.4,0.9) node{{\color{col4}$\blacksquare$}};
\draw (2.7,0.9) node{{\color{col4}$\blacksquare$}};
\draw (3,0.9) node{{\color{col4}$\blacksquare$}};
\draw (3.3,0.9) node{{\color{col4}$\blacksquare$}};
\draw (3.6,0.9) node{{\color{col4}$\blacksquare$}};
\draw (3.9,0.9) node{{\color{col4}$\blacksquare$}};
\draw (4.2,0.9) node{{\color{col4}$\blacksquare$}};
\end{scope}

\begin{scope}[xshift=5.2cm,yshift=-1.7cm]
\draw (0,0) node{{\color{col4}$\blacksquare$}};
\draw (0.3,0) node{{\color{col4}$\blacksquare$}};
\draw (0.6,0) node{{\color{col4}$\blacksquare$}};
\draw (0.9,0) node{{\color{col4}$\blacksquare$}};
\draw (1.2,0) node{{\color{col4}$\blacksquare$}};
\draw (1.5,0) node{{\color{col4}$\blacksquare$}};
\draw (1.8,0) node{{\color{col4}$\blacksquare$}};
\draw (2.1,0) node{{\color{col4}$\blacksquare$}};
\draw (2.4,0) node{{\color{col4}$\blacksquare$}};
\draw (2.7,0) node{{\color{col4}$\blacksquare$}};
\draw (3,0) node{{\color{col4}$\blacksquare$}};
\draw (3.3,0) node{{\color{col4}$\blacksquare$}};
\draw (3.6,0) node{{\color{col4}$\blacksquare$}};
\draw (3.9,0) node{{\color{col4}$\blacksquare$}};
\draw (4.2,0) node{{\color{col4}$\blacksquare$}};

\draw (0,0.3) node{{\color{col3}$\blacksquare$}};
\draw (0.3,0.3) node{{\color{col3}$\blacksquare$}};
\draw (0.6,0.3) node{{\color{col2}$\blacksquare$}};
\draw (0.9,0.3) node{{\color{col2}$\blacksquare$}};
\draw (1.2,0.3) node{{\color{col1}$\blacksquare$}};
\draw (1.5,0.3) node{{\color{col3}$\blacksquare$}};
\draw (1.8,0.3) node{{\color{col3}$\blacksquare$}};
\draw (2.1,0.3) node{{\color{col4}$\blacksquare$}};
\draw (2.4,0.3) node{{\color{col4}$\blacksquare$}};
\draw (2.7,0.3) node{{\color{col4}$\blacksquare$}};
\draw (3,0.3) node{{\color{col4}$\blacksquare$}};
\draw (3.3,0.3) node{{\color{col4}$\blacksquare$}};
\draw (3.6,0.3) node{{\color{col4}$\blacksquare$}};
\draw (3.9,0.3) node{{\color{col4}$\blacksquare$}};
\draw (4.2,0.3) node{{\color{col4}$\blacksquare$}};

\draw (0,0.6) node{{\color{col3}$\blacksquare$}};
\draw (0.3,0.6) node{{\color{col3}$\blacksquare$}};
\draw (0.6,0.6) node{{\color{col2}$\blacksquare$}};
\draw (0.9,0.6) node{{\color{col2}$\blacksquare$}};
\draw (1.2,0.6) node{{\color{col2}$\blacksquare$}};
\draw (1.5,0.6) node{{\color{col3}$\blacksquare$}};
\draw (1.8,0.6) node{{\color{col3}$\blacksquare$}};
\draw (2.1,0.6) node{{\color{col4}$\blacksquare$}};
\draw (2.4,0.6) node{{\color{col4}$\blacksquare$}};
\draw (2.7,0.6) node{{\color{col4}$\blacksquare$}};
\draw (3,0.6) node{{\color{col4}$\blacksquare$}};
\draw (3.3,0.6) node{{\color{col4}$\blacksquare$}};
\draw (3.6,0.6) node{{\color{col4}$\blacksquare$}};
\draw (3.9,0.6) node{{\color{col4}$\blacksquare$}};
\draw (4.2,0.6) node{{\color{col4}$\blacksquare$}};

\draw (0,0.9) node{{\color{col3}$\blacksquare$}};
\draw (0.3,0.9) node{{\color{col3}$\blacksquare$}};
\draw (0.6,0.9) node{{\color{col3}$\blacksquare$}};
\draw (0.9,0.9) node{{\color{col3}$\blacksquare$}};
\draw (1.2,0.9) node{{\color{col3}$\blacksquare$}};
\draw (1.5,0.9) node{{\color{col3}$\blacksquare$}};
\draw (1.8,0.9) node{{\color{col3}$\blacksquare$}};
\draw (2.1,0.9) node{{\color{col4}$\blacksquare$}};
\draw (2.4,0.9) node{{\color{col4}$\blacksquare$}};
\draw (2.7,0.9) node{{\color{col4}$\blacksquare$}};
\draw (3,0.9) node{{\color{col4}$\blacksquare$}};
\draw (3.3,0.9) node{{\color{col4}$\blacksquare$}};
\draw (3.6,0.9) node{{\color{col4}$\blacksquare$}};
\draw (3.9,0.9) node{{\color{col4}$\blacksquare$}};
\draw (4.2,0.9) node{{\color{col4}$\blacksquare$}};
\end{scope}

\begin{scope}[xshift=0cm,yshift=-3.4cm]
\draw (0,0) node{{\color{col4}$\blacksquare$}};
\draw (0.3,0) node{{\color{col4}$\blacksquare$}};
\draw (0.6,0) node{{\color{col4}$\blacksquare$}};
\draw (0.9,0) node{{\color{col4}$\blacksquare$}};
\draw (1.2,0) node{{\color{col4}$\blacksquare$}};
\draw (1.5,0) node{{\color{col4}$\blacksquare$}};
\draw (1.8,0) node{{\color{col4}$\blacksquare$}};
\draw (2.1,0) node{{\color{col4}$\blacksquare$}};
\draw (2.4,0) node{{\color{col4}$\blacksquare$}};
\draw (2.7,0) node{{\color{col4}$\blacksquare$}};
\draw (3,0) node{{\color{col4}$\blacksquare$}};
\draw (3.3,0) node{{\color{col4}$\blacksquare$}};
\draw (3.6,0) node{{\color{col4}$\blacksquare$}};
\draw (3.9,0) node{{\color{col4}$\blacksquare$}};
\draw (4.2,0) node{{\color{col4}$\blacksquare$}};

\draw (0,0.3) node{{\color{col3}$\blacksquare$}};
\draw (0.3,0.3) node{{\color{col3}$\blacksquare$}};
\draw (0.6,0.3) node{{\color{col3}$\blacksquare$}};
\draw (0.9,0.3) node{{\color{col3}$\blacksquare$}};
\draw (1.2,0.3) node{{\color{col1}$\blacksquare$}};
\draw (1.5,0.3) node{{\color{col2}$\blacksquare$}};
\draw (1.8,0.3) node{{\color{col2}$\blacksquare$}};
\draw (2.1,0.3) node{{\color{col4}$\blacksquare$}};
\draw (2.4,0.3) node{{\color{col4}$\blacksquare$}};
\draw (2.7,0.3) node{{\color{col4}$\blacksquare$}};
\draw (3,0.3) node{{\color{col4}$\blacksquare$}};
\draw (3.3,0.3) node{{\color{col4}$\blacksquare$}};
\draw (3.6,0.3) node{{\color{col4}$\blacksquare$}};
\draw (3.9,0.3) node{{\color{col4}$\blacksquare$}};
\draw (4.2,0.3) node{{\color{col4}$\blacksquare$}};

\draw (0,0.6) node{{\color{col3}$\blacksquare$}};
\draw (0.3,0.6) node{{\color{col3}$\blacksquare$}};
\draw (0.6,0.6) node{{\color{col3}$\blacksquare$}};
\draw (0.9,0.6) node{{\color{col3}$\blacksquare$}};
\draw (1.2,0.6) node{{\color{col2}$\blacksquare$}};
\draw (1.5,0.6) node{{\color{col2}$\blacksquare$}};
\draw (1.8,0.6) node{{\color{col2}$\blacksquare$}};
\draw (2.1,0.6) node{{\color{col4}$\blacksquare$}};
\draw (2.4,0.6) node{{\color{col4}$\blacksquare$}};
\draw (2.7,0.6) node{{\color{col4}$\blacksquare$}};
\draw (3,0.6) node{{\color{col4}$\blacksquare$}};
\draw (3.3,0.6) node{{\color{col4}$\blacksquare$}};
\draw (3.6,0.6) node{{\color{col4}$\blacksquare$}};
\draw (3.9,0.6) node{{\color{col4}$\blacksquare$}};
\draw (4.2,0.6) node{{\color{col4}$\blacksquare$}};

\draw (0,0.9) node{{\color{col3}$\blacksquare$}};
\draw (0.3,0.9) node{{\color{col3}$\blacksquare$}};
\draw (0.6,0.9) node{{\color{col3}$\blacksquare$}};
\draw (0.9,0.9) node{{\color{col3}$\blacksquare$}};
\draw (1.2,0.9) node{{\color{col3}$\blacksquare$}};
\draw (1.5,0.9) node{{\color{col3}$\blacksquare$}};
\draw (1.8,0.9) node{{\color{col3}$\blacksquare$}};
\draw (2.1,0.9) node{{\color{col4}$\blacksquare$}};
\draw (2.4,0.9) node{{\color{col4}$\blacksquare$}};
\draw (2.7,0.9) node{{\color{col4}$\blacksquare$}};
\draw (3,0.9) node{{\color{col4}$\blacksquare$}};
\draw (3.3,0.9) node{{\color{col4}$\blacksquare$}};
\draw (3.6,0.9) node{{\color{col4}$\blacksquare$}};
\draw (3.9,0.9) node{{\color{col4}$\blacksquare$}};
\draw (4.2,0.9) node{{\color{col4}$\blacksquare$}};
\end{scope}

\begin{scope}[xshift=5.2cm,yshift=-3.4cm]
\draw (0,0) node{{\color{col4}$\blacksquare$}};
\draw (0.3,0) node{{\color{col4}$\blacksquare$}};
\draw (0.6,0) node{{\color{col4}$\blacksquare$}};
\draw (0.9,0) node{{\color{col4}$\blacksquare$}};
\draw (1.2,0) node{{\color{col4}$\blacksquare$}};
\draw (1.5,0) node{{\color{col4}$\blacksquare$}};
\draw (1.8,0) node{{\color{col4}$\blacksquare$}};
\draw (2.1,0) node{{\color{col4}$\blacksquare$}};
\draw (2.4,0) node{{\color{col4}$\blacksquare$}};
\draw (2.7,0) node{{\color{col4}$\blacksquare$}};
\draw (3,0) node{{\color{col4}$\blacksquare$}};
\draw (3.3,0) node{{\color{col4}$\blacksquare$}};
\draw (3.6,0) node{{\color{col4}$\blacksquare$}};
\draw (3.9,0) node{{\color{col4}$\blacksquare$}};
\draw (4.2,0) node{{\color{col4}$\blacksquare$}};

\draw (0,0.3) node{{\color{col3}$\blacksquare$}};
\draw (0.3,0.3) node{{\color{col3}$\blacksquare$}};
\draw (0.6,0.3) node{{\color{col3}$\blacksquare$}};
\draw (0.9,0.3) node{{\color{col3}$\blacksquare$}};
\draw (1.2,0.3) node{{\color{col2}$\blacksquare$}};
\draw (1.5,0.3) node{{\color{col2}$\blacksquare$}};
\draw (1.8,0.3) node{{\color{col1}$\blacksquare$}};
\draw (2.1,0.3) node{{\color{col4}$\blacksquare$}};
\draw (2.4,0.3) node{{\color{col4}$\blacksquare$}};
\draw (2.7,0.3) node{{\color{col4}$\blacksquare$}};
\draw (3,0.3) node{{\color{col4}$\blacksquare$}};
\draw (3.3,0.3) node{{\color{col4}$\blacksquare$}};
\draw (3.6,0.3) node{{\color{col4}$\blacksquare$}};
\draw (3.9,0.3) node{{\color{col4}$\blacksquare$}};
\draw (4.2,0.3) node{{\color{col4}$\blacksquare$}};

\draw (0,0.6) node{{\color{col3}$\blacksquare$}};
\draw (0.3,0.6) node{{\color{col3}$\blacksquare$}};
\draw (0.6,0.6) node{{\color{col3}$\blacksquare$}};
\draw (0.9,0.6) node{{\color{col3}$\blacksquare$}};
\draw (1.2,0.6) node{{\color{col2}$\blacksquare$}};
\draw (1.5,0.6) node{{\color{col2}$\blacksquare$}};
\draw (1.8,0.6) node{{\color{col2}$\blacksquare$}};
\draw (2.1,0.6) node{{\color{col4}$\blacksquare$}};
\draw (2.4,0.6) node{{\color{col4}$\blacksquare$}};
\draw (2.7,0.6) node{{\color{col4}$\blacksquare$}};
\draw (3,0.6) node{{\color{col4}$\blacksquare$}};
\draw (3.3,0.6) node{{\color{col4}$\blacksquare$}};
\draw (3.6,0.6) node{{\color{col4}$\blacksquare$}};
\draw (3.9,0.6) node{{\color{col4}$\blacksquare$}};
\draw (4.2,0.6) node{{\color{col4}$\blacksquare$}};

\draw (0,0.9) node{{\color{col3}$\blacksquare$}};
\draw (0.3,0.9) node{{\color{col3}$\blacksquare$}};
\draw (0.6,0.9) node{{\color{col3}$\blacksquare$}};
\draw (0.9,0.9) node{{\color{col3}$\blacksquare$}};
\draw (1.2,0.9) node{{\color{col3}$\blacksquare$}};
\draw (1.5,0.9) node{{\color{col3}$\blacksquare$}};
\draw (1.8,0.9) node{{\color{col3}$\blacksquare$}};
\draw (2.1,0.9) node{{\color{col4}$\blacksquare$}};
\draw (2.4,0.9) node{{\color{col4}$\blacksquare$}};
\draw (2.7,0.9) node{{\color{col4}$\blacksquare$}};
\draw (3,0.9) node{{\color{col4}$\blacksquare$}};
\draw (3.3,0.9) node{{\color{col4}$\blacksquare$}};
\draw (3.6,0.9) node{{\color{col4}$\blacksquare$}};
\draw (3.9,0.9) node{{\color{col4}$\blacksquare$}};
\draw (4.2,0.9) node{{\color{col4}$\blacksquare$}};
\end{scope}

\begin{scope}[xshift=0cm,yshift=-5.1cm]
\draw (0,0) node{{\color{col4}$\blacksquare$}};
\draw (0.3,0) node{{\color{col4}$\blacksquare$}};
\draw (0.6,0) node{{\color{col4}$\blacksquare$}};
\draw (0.9,0) node{{\color{col4}$\blacksquare$}};
\draw (1.2,0) node{{\color{col3}$\blacksquare$}};
\draw (1.5,0) node{{\color{col3}$\blacksquare$}};
\draw (1.8,0) node{{\color{col3}$\blacksquare$}};
\draw (2.1,0) node{{\color{col3}$\blacksquare$}};
\draw (2.4,0) node{{\color{col3}$\blacksquare$}};
\draw (2.7,0) node{{\color{col3}$\blacksquare$}};
\draw (3,0) node{{\color{col3}$\blacksquare$}};
\draw (3.3,0) node{{\color{col4}$\blacksquare$}};
\draw (3.6,0) node{{\color{col4}$\blacksquare$}};
\draw (3.9,0) node{{\color{col4}$\blacksquare$}};
\draw (4.2,0) node{{\color{col4}$\blacksquare$}};

\draw (0,0.3) node{{\color{col4}$\blacksquare$}};
\draw (0.3,0.3) node{{\color{col4}$\blacksquare$}};
\draw (0.6,0.3) node{{\color{col4}$\blacksquare$}};
\draw (0.9,0.3) node{{\color{col4}$\blacksquare$}};
\draw (1.2,0.3) node{{\color{col2}$\blacksquare$}};
\draw (1.5,0.3) node{{\color{col2}$\blacksquare$}};
\draw (1.8,0.3) node{{\color{col1}$\blacksquare$}};
\draw (2.1,0.3) node{{\color{col3}$\blacksquare$}};
\draw (2.4,0.3) node{{\color{col3}$\blacksquare$}};
\draw (2.7,0.3) node{{\color{col3}$\blacksquare$}};
\draw (3,0.3) node{{\color{col3}$\blacksquare$}};
\draw (3.3,0.3) node{{\color{col4}$\blacksquare$}};
\draw (3.6,0.3) node{{\color{col4}$\blacksquare$}};
\draw (3.9,0.3) node{{\color{col4}$\blacksquare$}};
\draw (4.2,0.3) node{{\color{col4}$\blacksquare$}};

\draw (0,0.6) node{{\color{col4}$\blacksquare$}};
\draw (0.3,0.6) node{{\color{col4}$\blacksquare$}};
\draw (0.6,0.6) node{{\color{col4}$\blacksquare$}};
\draw (0.9,0.6) node{{\color{col4}$\blacksquare$}};
\draw (1.2,0.6) node{{\color{col2}$\blacksquare$}};
\draw (1.5,0.6) node{{\color{col2}$\blacksquare$}};
\draw (1.8,0.6) node{{\color{col2}$\blacksquare$}};
\draw (2.1,0.6) node{{\color{col3}$\blacksquare$}};
\draw (2.4,0.6) node{{\color{col3}$\blacksquare$}};
\draw (2.7,0.6) node{{\color{col3}$\blacksquare$}};
\draw (3,0.6) node{{\color{col3}$\blacksquare$}};
\draw (3.3,0.6) node{{\color{col4}$\blacksquare$}};
\draw (3.6,0.6) node{{\color{col4}$\blacksquare$}};
\draw (3.9,0.6) node{{\color{col4}$\blacksquare$}};
\draw (4.2,0.6) node{{\color{col4}$\blacksquare$}};

\draw (0,0.9) node{{\color{col4}$\blacksquare$}};
\draw (0.3,0.9) node{{\color{col4}$\blacksquare$}};
\draw (0.6,0.9) node{{\color{col4}$\blacksquare$}};
\draw (0.9,0.9) node{{\color{col4}$\blacksquare$}};
\draw (1.2,0.9) node{{\color{col4}$\blacksquare$}};
\draw (1.5,0.9) node{{\color{col4}$\blacksquare$}};
\draw (1.8,0.9) node{{\color{col4}$\blacksquare$}};
\draw (2.1,0.9) node{{\color{col4}$\blacksquare$}};
\draw (2.4,0.9) node{{\color{col4}$\blacksquare$}};
\draw (2.7,0.9) node{{\color{col4}$\blacksquare$}};
\draw (3,0.9) node{{\color{col4}$\blacksquare$}};
\draw (3.3,0.9) node{{\color{col4}$\blacksquare$}};
\draw (3.6,0.9) node{{\color{col4}$\blacksquare$}};
\draw (3.9,0.9) node{{\color{col4}$\blacksquare$}};
\draw (4.2,0.9) node{{\color{col4}$\blacksquare$}};
\end{scope}

\begin{scope}[xshift=5.2cm,yshift=-5.1cm]
\draw (0,0) node{{\color{col4}$\blacksquare$}};
\draw (0.3,0) node{{\color{col4}$\blacksquare$}};
\draw (0.6,0) node{{\color{col4}$\blacksquare$}};
\draw (0.9,0) node{{\color{col4}$\blacksquare$}};
\draw (1.2,0) node{{\color{col3}$\blacksquare$}};
\draw (1.5,0) node{{\color{col3}$\blacksquare$}};
\draw (1.8,0) node{{\color{col3}$\blacksquare$}};
\draw (2.1,0) node{{\color{col3}$\blacksquare$}};
\draw (2.4,0) node{{\color{col2}$\blacksquare$}};
\draw (2.7,0) node{{\color{col2}$\blacksquare$}};
\draw (3,0) node{{\color{col1}$\blacksquare$}};
\draw (3.3,0) node{{\color{col4}$\blacksquare$}};
\draw (3.6,0) node{{\color{col4}$\blacksquare$}};
\draw (3.9,0) node{{\color{col4}$\blacksquare$}};
\draw (4.2,0) node{{\color{col4}$\blacksquare$}};

\draw (0,0.3) node{{\color{col4}$\blacksquare$}};
\draw (0.3,0.3) node{{\color{col4}$\blacksquare$}};
\draw (0.6,0.3) node{{\color{col4}$\blacksquare$}};
\draw (0.9,0.3) node{{\color{col4}$\blacksquare$}};
\draw (1.2,0.3) node{{\color{col3}$\blacksquare$}};
\draw (1.5,0.3) node{{\color{col3}$\blacksquare$}};
\draw (1.8,0.3) node{{\color{col3}$\blacksquare$}};
\draw (2.1,0.3) node{{\color{col3}$\blacksquare$}};
\draw (2.4,0.3) node{{\color{col2}$\blacksquare$}};
\draw (2.7,0.3) node{{\color{col2}$\blacksquare$}};
\draw (3,0.3) node{{\color{col2}$\blacksquare$}};
\draw (3.3,0.3) node{{\color{col4}$\blacksquare$}};
\draw (3.6,0.3) node{{\color{col4}$\blacksquare$}};
\draw (3.9,0.3) node{{\color{col4}$\blacksquare$}};
\draw (4.2,0.3) node{{\color{col4}$\blacksquare$}};

\draw (0,0.6) node{{\color{col4}$\blacksquare$}};
\draw (0.3,0.6) node{{\color{col4}$\blacksquare$}};
\draw (0.6,0.6) node{{\color{col4}$\blacksquare$}};
\draw (0.9,0.6) node{{\color{col4}$\blacksquare$}};
\draw (1.2,0.6) node{{\color{col3}$\blacksquare$}};
\draw (1.5,0.6) node{{\color{col3}$\blacksquare$}};
\draw (1.8,0.6) node{{\color{col3}$\blacksquare$}};
\draw (2.1,0.6) node{{\color{col3}$\blacksquare$}};
\draw (2.4,0.6) node{{\color{col3}$\blacksquare$}};
\draw (2.7,0.6) node{{\color{col3}$\blacksquare$}};
\draw (3,0.6) node{{\color{col3}$\blacksquare$}};
\draw (3.3,0.6) node{{\color{col4}$\blacksquare$}};
\draw (3.6,0.6) node{{\color{col4}$\blacksquare$}};
\draw (3.9,0.6) node{{\color{col4}$\blacksquare$}};
\draw (4.2,0.6) node{{\color{col4}$\blacksquare$}};

\draw (0,0.9) node{{\color{col4}$\blacksquare$}};
\draw (0.3,0.9) node{{\color{col4}$\blacksquare$}};
\draw (0.6,0.9) node{{\color{col4}$\blacksquare$}};
\draw (0.9,0.9) node{{\color{col4}$\blacksquare$}};
\draw (1.2,0.9) node{{\color{col4}$\blacksquare$}};
\draw (1.5,0.9) node{{\color{col4}$\blacksquare$}};
\draw (1.8,0.9) node{{\color{col4}$\blacksquare$}};
\draw (2.1,0.9) node{{\color{col4}$\blacksquare$}};
\draw (2.4,0.9) node{{\color{col4}$\blacksquare$}};
\draw (2.7,0.9) node{{\color{col4}$\blacksquare$}};
\draw (3,0.9) node{{\color{col4}$\blacksquare$}};
\draw (3.3,0.9) node{{\color{col4}$\blacksquare$}};
\draw (3.6,0.9) node{{\color{col4}$\blacksquare$}};
\draw (3.9,0.9) node{{\color{col4}$\blacksquare$}};
\draw (4.2,0.9) node{{\color{col4}$\blacksquare$}};
\end{scope}

\begin{scope}[xshift=0cm,yshift=-6.8cm]
\draw (0,0) node{{\color{col4}$\blacksquare$}};
\draw (0.3,0) node{{\color{col4}$\blacksquare$}};
\draw (0.6,0) node{{\color{col4}$\blacksquare$}};
\draw (0.9,0) node{{\color{col4}$\blacksquare$}};
\draw (1.2,0) node{{\color{col4}$\blacksquare$}};
\draw (1.5,0) node{{\color{col4}$\blacksquare$}};
\draw (1.8,0) node{{\color{col4}$\blacksquare$}};
\draw (2.1,0) node{{\color{col4}$\blacksquare$}};
\draw (2.4,0) node{{\color{col2}$\blacksquare$}};
\draw (2.7,0) node{{\color{col2}$\blacksquare$}};
\draw (3,0) node{{\color{col1}$\blacksquare$}};
\draw (3.3,0) node{{\color{col3}$\blacksquare$}};
\draw (3.6,0) node{{\color{col3}$\blacksquare$}};
\draw (3.9,0) node{{\color{col3}$\blacksquare$}};
\draw (4.2,0) node{{\color{col3}$\blacksquare$}};

\draw (0,0.3) node{{\color{col4}$\blacksquare$}};
\draw (0.3,0.3) node{{\color{col4}$\blacksquare$}};
\draw (0.6,0.3) node{{\color{col4}$\blacksquare$}};
\draw (0.9,0.3) node{{\color{col4}$\blacksquare$}};
\draw (1.2,0.3) node{{\color{col4}$\blacksquare$}};
\draw (1.5,0.3) node{{\color{col4}$\blacksquare$}};
\draw (1.8,0.3) node{{\color{col4}$\blacksquare$}};
\draw (2.1,0.3) node{{\color{col4}$\blacksquare$}};
\draw (2.4,0.3) node{{\color{col2}$\blacksquare$}};
\draw (2.7,0.3) node{{\color{col2}$\blacksquare$}};
\draw (3,0.3) node{{\color{col2}$\blacksquare$}};
\draw (3.3,0.3) node{{\color{col3}$\blacksquare$}};
\draw (3.6,0.3) node{{\color{col3}$\blacksquare$}};
\draw (3.9,0.3) node{{\color{col3}$\blacksquare$}};
\draw (4.2,0.3) node{{\color{col3}$\blacksquare$}};

\draw (0,0.6) node{{\color{col4}$\blacksquare$}};
\draw (0.3,0.6) node{{\color{col4}$\blacksquare$}};
\draw (0.6,0.6) node{{\color{col4}$\blacksquare$}};
\draw (0.9,0.6) node{{\color{col4}$\blacksquare$}};
\draw (1.2,0.6) node{{\color{col4}$\blacksquare$}};
\draw (1.5,0.6) node{{\color{col4}$\blacksquare$}};
\draw (1.8,0.6) node{{\color{col4}$\blacksquare$}};
\draw (2.1,0.6) node{{\color{col4}$\blacksquare$}};
\draw (2.4,0.6) node{{\color{col3}$\blacksquare$}};
\draw (2.7,0.6) node{{\color{col3}$\blacksquare$}};
\draw (3,0.6) node{{\color{col3}$\blacksquare$}};
\draw (3.3,0.6) node{{\color{col3}$\blacksquare$}};
\draw (3.6,0.6) node{{\color{col3}$\blacksquare$}};
\draw (3.9,0.6) node{{\color{col3}$\blacksquare$}};
\draw (4.2,0.6) node{{\color{col3}$\blacksquare$}};

\draw (0,0.9) node{{\color{col4}$\blacksquare$}};
\draw (0.3,0.9) node{{\color{col4}$\blacksquare$}};
\draw (0.6,0.9) node{{\color{col4}$\blacksquare$}};
\draw (0.9,0.9) node{{\color{col4}$\blacksquare$}};
\draw (1.2,0.9) node{{\color{col4}$\blacksquare$}};
\draw (1.5,0.9) node{{\color{col4}$\blacksquare$}};
\draw (1.8,0.9) node{{\color{col4}$\blacksquare$}};
\draw (2.1,0.9) node{{\color{col4}$\blacksquare$}};
\draw (2.4,0.9) node{{\color{col4}$\blacksquare$}};
\draw (2.7,0.9) node{{\color{col4}$\blacksquare$}};
\draw (3,0.9) node{{\color{col4}$\blacksquare$}};
\draw (3.3,0.9) node{{\color{col4}$\blacksquare$}};
\draw (3.6,0.9) node{{\color{col4}$\blacksquare$}};
\draw (3.9,0.9) node{{\color{col4}$\blacksquare$}};
\draw (4.2,0.9) node{{\color{col4}$\blacksquare$}};
\end{scope}

\begin{scope}[xshift=5.2cm,yshift=-6.8cm]
\draw (0,0) node{{\color{col4}$\blacksquare$}};
\draw (0.3,0) node{{\color{col4}$\blacksquare$}};
\draw (0.6,0) node{{\color{col4}$\blacksquare$}};
\draw (0.9,0) node{{\color{col4}$\blacksquare$}};
\draw (1.2,0) node{{\color{col4}$\blacksquare$}};
\draw (1.5,0) node{{\color{col4}$\blacksquare$}};
\draw (1.8,0) node{{\color{col4}$\blacksquare$}};
\draw (2.1,0) node{{\color{col4}$\blacksquare$}};
\draw (2.4,0) node{{\color{col3}$\blacksquare$}};
\draw (2.7,0) node{{\color{col3}$\blacksquare$}};
\draw (3,0) node{{\color{col3}$\blacksquare$}};
\draw (3.3,0) node{{\color{col3}$\blacksquare$}};
\draw (3.6,0) node{{\color{col2}$\blacksquare$}};
\draw (3.9,0) node{{\color{col2}$\blacksquare$}};
\draw (4.2,0) node{{\color{col1}$\blacksquare$}};

\draw (0,0.3) node{{\color{col4}$\blacksquare$}};
\draw (0.3,0.3) node{{\color{col4}$\blacksquare$}};
\draw (0.6,0.3) node{{\color{col4}$\blacksquare$}};
\draw (0.9,0.3) node{{\color{col4}$\blacksquare$}};
\draw (1.2,0.3) node{{\color{col4}$\blacksquare$}};
\draw (1.5,0.3) node{{\color{col4}$\blacksquare$}};
\draw (1.8,0.3) node{{\color{col4}$\blacksquare$}};
\draw (2.1,0.3) node{{\color{col4}$\blacksquare$}};
\draw (2.4,0.3) node{{\color{col3}$\blacksquare$}};
\draw (2.7,0.3) node{{\color{col3}$\blacksquare$}};
\draw (3,0.3) node{{\color{col3}$\blacksquare$}};
\draw (3.3,0.3) node{{\color{col3}$\blacksquare$}};
\draw (3.6,0.3) node{{\color{col2}$\blacksquare$}};
\draw (3.9,0.3) node{{\color{col2}$\blacksquare$}};
\draw (4.2,0.3) node{{\color{col2}$\blacksquare$}};

\draw (0,0.6) node{{\color{col4}$\blacksquare$}};
\draw (0.3,0.6) node{{\color{col4}$\blacksquare$}};
\draw (0.6,0.6) node{{\color{col4}$\blacksquare$}};
\draw (0.9,0.6) node{{\color{col4}$\blacksquare$}};
\draw (1.2,0.6) node{{\color{col4}$\blacksquare$}};
\draw (1.5,0.6) node{{\color{col4}$\blacksquare$}};
\draw (1.8,0.6) node{{\color{col4}$\blacksquare$}};
\draw (2.1,0.6) node{{\color{col4}$\blacksquare$}};
\draw (2.4,0.6) node{{\color{col3}$\blacksquare$}};
\draw (2.7,0.6) node{{\color{col3}$\blacksquare$}};
\draw (3,0.6) node{{\color{col3}$\blacksquare$}};
\draw (3.3,0.6) node{{\color{col3}$\blacksquare$}};
\draw (3.6,0.6) node{{\color{col3}$\blacksquare$}};
\draw (3.9,0.6) node{{\color{col3}$\blacksquare$}};
\draw (4.2,0.6) node{{\color{col3}$\blacksquare$}};

\draw (0,0.9) node{{\color{col4}$\blacksquare$}};
\draw (0.3,0.9) node{{\color{col4}$\blacksquare$}};
\draw (0.6,0.9) node{{\color{col4}$\blacksquare$}};
\draw (0.9,0.9) node{{\color{col4}$\blacksquare$}};
\draw (1.2,0.9) node{{\color{col4}$\blacksquare$}};
\draw (1.5,0.9) node{{\color{col4}$\blacksquare$}};
\draw (1.8,0.9) node{{\color{col4}$\blacksquare$}};
\draw (2.1,0.9) node{{\color{col4}$\blacksquare$}};
\draw (2.4,0.9) node{{\color{col4}$\blacksquare$}};
\draw (2.7,0.9) node{{\color{col4}$\blacksquare$}};
\draw (3,0.9) node{{\color{col4}$\blacksquare$}};
\draw (3.3,0.9) node{{\color{col4}$\blacksquare$}};
\draw (3.6,0.9) node{{\color{col4}$\blacksquare$}};
\draw (3.9,0.9) node{{\color{col4}$\blacksquare$}};
\draw (4.2,0.9) node{{\color{col4}$\blacksquare$}};
\end{scope}

\end{tikzpicture}
\end{center}

\caption{
\label{figure: flemish mountain}
Possible motion of particles inside a bubble. 
${\color{col1}\blacksquare}$: site with 3 particles, 
${\color{col2}\blacksquare}$: site with 2 particles,
${\color{col3}\blacksquare}$: site with 1 particle,
${\color{col4}\blacksquare}$: site with no particle.
From left to right and from top to bottom: 
as an example, we use a systematic procedure to show that a classical state where the site occupied by 3 particles is in the upper left corner, 
is connected via $H_{res}$ to a classical state where this site is in the lower right corner. 
}
\end{figure}

\section{Rare bubbles due to non-perturbative effects}\label{section: bubbles non perturbative}

A drawback of the analysis of Section \ref{rare bubbles perturbative} is that it relies on fine details of the model, there given by \eqref{Bose Hubbard d=2}. 
It is actually shown in \cite{3}, and we will see another example below, that if only nearest neighbor hopping was present in \eqref{Bose Hubbard Hamiltonian}, 
all but a vanishing proportion of the eigenstates of the corresponding first order resonant Hamiltonian would be localized. 
It could thus seem conceivable that, by restricting our attention to a well-chosen class of models, 
we cannot find any delocalizing bubble at any order, and hence conclude the existence of a localized phase.

We aim to show here that this is not the case, once non-perturbative effects are taken into account. 
Once again, we follow \cite{3} (a similar reasoning has been developed by \cite{Huse Nandkishore private}).
While in the previous section we proceeded via a detailed analysis at low orders, 
we now assume that, for the full dynamics, ETH is satisfied inside low density bubbles. 
When their size is large enough, these bubbles can act as baths: they absorb and eject particles from and to the system. 
It is worth pointing out that such baths are eventually observed in quenched disordered systems too, such as \eqref{Hamiltonian disordered chain}, due to large deviations in the disorder that appear unavoidably at large enough volume, no matter how strong the disorder is.
In both cases, delocalization does not follow at once since, due to their finite size, these baths are not perfect. 
The difference between quenched disordered and translation invariant systems comes once again from the fact that, 
in the latter case, the bubbles can be located everywhere and can therefore move across the system by interacting with the rest of the system, 
while in the former case, their location can be determined a priori if the disorder is strong enough. 

We now proceed to a more precise description.
Though we study a rather specific case, where our argument can be stated neatly, the described phenomenon is very general.
In particular, we expect it to show up for all the Hamiltonians of the type (\ref{general sum of local terms Hamiltonian}-\ref{general perturbative Hamiltonian}).
In contrast to the cases considered until now, in models \eqref{Bose Hubbard Hamiltonian} and \eqref{Bose Hubbard d=2},
we do not longer take $H^{(0)}$ to be uniformly localized.
We know indeed that in these examples, $H$ cannot be such, and we wish to incorporate this feature into $H^{(0)}$. 
Below, we first describe $H^{(0)}$ in more details, and then show how arbitrarily small perturbations force all eigenstates of $H^{(0)}$ to hybridize to form the eigenstates of $H$.

\paragraph{The Hamiltonian $H^{(0)}$.}
We let $d=2$ and $\NumLevels \gg 1$. 
The Hamiltonian $H^{(0)}$ is itself the sum of a diagonal interaction term and a hopping term: $H^{(0)} = H^{(0)}_i + \lambda H^{(0)}_h$. 
We assume $\coupling \ll \lambda \ll 1$.
A natural choice for $H^{(0)}_h$ would be given by $H^{(0)}_h = \frac{1}{2}\sum_{x\sim y} \{ b_x^* b_y + b_x b_y^* \}$ but,
since non-resonant transitions can be removed perturbatively at first order in $\lambda$ as pointed out in the previous section, 
let us consider a caricature, similar to \eqref{resonant Bose Hubbard d=2}, where only resonant hoppings appear: 
\begin{multline}\label{H 0 bubble model}
\langle \eta' | H^{(0)} | \eta \rangle 
\; = \; \\
\langle \eta' | H^{(0)}_i | \eta \rangle \, + \,  
\frac{\lambda}{2} \, \delta\left( H^{(0)}_i (\eta') - H^{(0)}_i (\eta) \right) \sum_{x,y:x \sim y}  \langle \eta' | b_x^* b_y + b_x b_y^* | \eta\rangle .
\end{multline}

Let us specify the interaction energy $H_i^{(0)}$. We set
\begin{equation}\label{H 0 i bubble model}
H^{(0)}_i = \sum_{A \subset \Z^2: |A| = 2} H_A^{(0)} \qquad\quad (A \text{ connected}). 
\end{equation}
The operators $H_A^{(0)}$ are diagonal in the $\{|\eta \rangle \}$ basis, and are such that, for any translation $\tau_x$ by $x$, 
it holds that $H^{(0)}_{\tau_x A} (\eta)  = H^{(0)}_A ( \tau_{-x} \eta )$, so that $H^{(0)}_i$ is translation invariant. 
Therefore, we only need to specify $H^{(0)}_A$ for $A$ being  $A_1 = \{ (0,0),(1,0)\}$ or $A_2 = \{ (0,0),(0,1)\}$, 
so two functions $f_1,f_2 : \{ 0, \dots , \NumLevels \}^2 \to \R$ such that 
\begin{equation*}
H_{A_1}^{(0)} ( \eta ) \; = \; f_1 \left(\eta_{(0,0)}, \eta_{(1,0)}\right), 
\qquad 
H_{A_2}^{(0)} ( \eta ) \; = \; f_2 \left(\eta_{(0,0)}, \eta_{(0,1)}\right).
\end{equation*}
In anlogy with the Bose-Hubbard Hamiltonian, we require that typically $H_{A_j}^{(0)}(\eta) \sim \NumLevels^2$ ($j=1,2$).
Moreover, we impose a genericity condition on $f_1, f_2$: 
these functions are injective, the set of values of $f_1$ has no intersection with the set of values of $f_2$, 
and the values $\alpha_1, \dots , \alpha_{2 (\NumLevels+1)^2}\in f_1 (\{0,\dots N\}^2) \cup f_2 (\{0,\dots N\}^2)$ are not in any rational combination of each others: 
\begin{equation}\label{genericity condition}
k_1 \alpha_1 + \dots + k_{2 (\NumLevels+1)^2} \alpha_{2 (\NumLevels+1)^2} \ne 0 
\qquad \forall \; 
(k_1, \dots, k_{2 (\NumLevels+1)^2}) \in \Z^{2 (\NumLevels+1)^2} \backslash \{0\}.
\end{equation}
We can think of the values $\alpha_1, \dots , \alpha_{2 (\NumLevels+1)^2}$ as $2(\NumLevels+1)^2$ numbers picked at random in an interval of size 1. 
In particular, $H_i^{(0)}$ is not invariant under lattice rotations. 

The study of the localized and delocalized phases of the Hamiltonian $H^{(0)}$ defined through (\ref{H 0 bubble model}-\ref{H 0 i bubble model}) 
boils down to a percolation analysis in $\Z^2$.  
This is a consequence of the following property, shown in Appendix 2. 
Let $|\eta \rangle$ be an initial classical state, and let $| \phi (t) \rangle = \ed^{-it H^{(0)}} | \eta \rangle$ be its time evolution after a time $t \ge 0$. 
Consider four distinct points $x,y,z,w$ forming a square of nearest neighbors: 
\begin{equation*}
|x-y|_1 \; = \; |y-z|_1 \; = \; |z-w|_1 \; = \; |w-x|_1 \; = \; 1.
\end{equation*}
If it holds that 
\begin{equation}\label{carre de lu}
|\eta_x - \eta_y| \; \ge \; 2, \quad
|\eta_y - \eta_z| \; \ge \; 2, \quad
|\eta_z - \eta_w| \; \ge \; 2, \quad
|\eta_w - \eta_x| \; \ge \; 2, 
\end{equation}
then, for all $t\ge 0$, it holds that 
\begin{equation}\label{stability carre de lu}
\phi_x(t) = \eta_x, \quad
\phi_y(t) = \eta_y, \quad
\phi_z(t) = \eta_z, \quad
\phi_w(t) = \eta_w.
\end{equation}

The following picture emerges. 
At very low density, for a typical initial classical configuration $| \eta \rangle$, particles are few and far between.
Under the dynamics generated by $H^{(0)}$, given by (\ref{H 0 bubble model}-\ref{H 0 i bubble model}), isolated particles evolve like on a billiard: they move freely with the additional constraint that two particles can never come at a distance less or equal to one from each other.
Therefore, though a cluster of frozen particles appears here and there, for example because some particles are in a configuration satisfying (\ref{carre de lu}-\ref{stability carre de lu}), 
we observe a low density gas of interacting particles. 
Ergodic behavior is expected. 

At higher density instead, above some threshold, 
for a typical initial classical state $| \eta \rangle$, the sites included in a square of nearest neighbors where (\ref{carre de lu}-\ref{stability carre de lu}) is satisfied, 
form a giant cluster in the lattice, so that the dynamics is frozen. 
The system is in the MBL phase. 
Nevertheless, the localization length is not uniform: it becomes arbitrarily large on some islands, whose locations are determined by the initial state $| \eta \rangle$.  
Indeed, in the thermodynamic limit, low density bubbles of arbitrary large size are found in a typical classical configuration. 
Since normal thermal behavior is again expected inside the bubbles, the localization length becomes there of the size of the bubble.

In view of the above, despite the rather specific, and perhaps artificial, definition of $H^{(0)}$ given by (\ref{H 0 bubble model}-\ref{H 0 i bubble model}), 
we hope that this model furnishes a reasonable cartoon of the MBL phase in translation invariant systems, 
so that our conclusions will ultimately not depend on the particular choice of this model.

\paragraph{Hybridization of eigenstates.}
Let us assume that the density is large enough so that $H^{(0)}$ is in the localized phase described above. 
Moreover, we assume that $H^{(1)}$ allows for nearest and second neighbor hopping. 
Let us give two classical states $|\eta\rangle$ and $| \eta' \rangle$. 
To simplify the discussion let us assume that the volume $V$ is large but finite, and that $|\eta \rangle$ and $| \eta' \rangle$ contain only one thermal bubble, located at precisely the same place, while all the other sites are frozen.
The state $| \eta' \rangle$ is distinct from $|\eta \rangle$ in that a single particle on a frozen site on the outside border of the bubble in $|\eta \rangle$ has moved inside the bubble in $| \eta' \rangle$; we have $\langle \eta' | H^{(0)} | \eta \rangle = 0$ but $\langle \eta' | H^{(1)} | \eta \rangle \ne 0$.
We denote by $| \psi \rangle $ any eigenstate of $H^{(0)}$ such that $\langle \eta | \psi \rangle \ne 0$, 
and by $| \psi' \rangle $ any eigenstate of $H^{(0)}$ such that $\langle \eta' | \psi' \rangle \ne 0$.

We show below that, as soon as the size of the bubble is big enough,
given any state $| \psi \rangle$, we can find at least one (but actually much more than one) state $| \psi' \rangle$ such that 
\begin{equation}\label{resonant transition bubble}
\left|\frac{\coupling \langle \psi' | H^{(1)} | \psi \rangle }{ H^{(0)} ( \psi' ) - H^{(0)} ( \psi ) } \right| \; \gg \; 1. 
\end{equation}
This means that the extraction of a particle from the surrounding to the inside of the bubble can be achieved via a resonant transition. 
As this is then true for the reverse process too, we conclude, as in the previous section, 
that the bubble can eventually move everywhere by absorbing and expelling particles. 

In \cite{3}, \eqref{resonant transition bubble} is shown to follow from ETH inside the bubble. 
In the hope of being more explicit, we here prove \eqref{resonant transition bubble} under stronger hypotheses (Berry conjecture, see \cite{Srednicki}).
Let us denote by $\mathcal N$ the dimension of the Hilbert space spanned by all the calssical states that are connected to $|\eta \rangle$ via a sequence of resonant transitions (see \eqref{connected via Hamiltonian}), and by $|\mathcal B|$ the physical size of the bubble. 
It holds that $\mathcal N \sim \ed^{s |\mathcal B|}$ for some entropy density $s > 0$, that depends on the density of particles inside the bubble. 
Obviously, the values $\mathcal N$ and $\mathcal B$ would be slightly different for $| \eta\rangle$ replaced by $| \eta' \rangle$ (i.e.\@ the values change slightly before and after the absorption of a particle from the surrounding), but we neglect this as it is irrelevant for the present computation. 
Let us fix $| \psi \rangle$. 

We first estimate the numerator in \eqref{resonant transition bubble},
\begin{equation*}
\langle \psi' | H^{(1)} | \psi \rangle 
\; = \;
\sum_{\tilde\eta,\tilde\eta'} \langle \psi' | \tilde \eta' \rangle \langle \tilde \eta | \psi \rangle \langle \tilde \eta' | H^{(1)} | \tilde\eta \rangle,   
\end{equation*}
where the sum ranges over the classical states $\tilde\eta$ and $\tilde\eta'$ such that $\langle \tilde\eta | \ed^{-i t H^{(0)} } | \eta \rangle \ne 0$ for some $t\ge 0$, 
and $\langle \tilde\eta' | \ed^{-i t' H^{(0)} } | \eta' \rangle \ne 0$ for some $t'\ge 0$
(the sum is thus over $\mathcal N \times \mathcal N$ states). 
Since the bubble is ergodic, we assume, inspired by \cite{Srednicki}, that
\begin{equation}\label{ETH inside the bubble}
\langle \psi' | \tilde \eta' \rangle \; \sim \; \frac{\ed^{i \theta(\psi',\tilde\eta')}}{\sqrt\mathcal N}, 
\qquad 
\langle \eta | \tilde \psi \rangle \; \sim \; \frac{\ed^{-i \theta(\psi,\tilde\eta)}}{\sqrt\mathcal N}
\end{equation}
for some phases $\theta(\psi',\tilde\eta')$ and $ \theta(\psi,\tilde\eta)$. 
Next, for any state $|\tilde\eta \rangle$, there are at most $|\mathcal B|$ states $| \tilde \eta' \rangle$ such that $\langle \tilde \eta' | H^{(1)} | \tilde\eta \rangle \ne 0$.
Moreover, following again \cite{Srednicki}, we assume that the phases in \eqref{ETH inside the bubble} are distributed randomly.
Therefore, by a central limit argument, we find
\begin{equation*}
\coupling \left| \langle \psi' | H^{(1)} | \psi \rangle \right|  
\; \lesssim \;  
\coupling \frac{|\mathcal B|\sqrt \mathcal N}{\mathcal N} 
\; = \; 
\frac{\coupling |\mathcal B|}{ \sqrt{\mathcal N}}. 
\end{equation*}

Let us then look at the denominator in \eqref{resonant transition bubble}. 
Since there are $\mathcal N$ possible states $| \psi' \rangle$, the denominator takes $\mathcal N$ possible values depending on the choice of $|\psi' \rangle$, 
all sitting in an interval of length $|\mathcal B|$. 
Assuming them to be approximately equidistributed, we conclude that, for some $| \psi' \rangle$, the denominator is of size $|\mathcal B|/\mathcal N$. 
Altogether we conclude that
\begin{equation*}
\left|\frac{\coupling \langle \psi' | H^{(1)} | \psi \rangle }{H^{(0)} ( \psi' ) - H^{(0)} ( \psi ) } \right| 
\; \sim \;
\frac{\coupling \mathcal N }{\sqrt{\mathcal N}} 
\; \gg \; 1
\end{equation*}
for a large enough bubble.
This is \eqref{resonant transition bubble}.

\section{Asymptotic localization}\label{section: asymptotic localization}

For this last section, we consider both classical and quantum systems, and we restrict ourselves to $d=1$ though this is probably not at all necessary. 
We assume that $V = \{ - (|V|-1)/2, \dots, (|V|-1)/2 \}$ for some odd integer $|V|\ge 1$. 
We still take a Hamiltonian of the form \eqref{general sum of local terms Hamiltonian}, but write it now more simply as $H = \sum_{x\in V} H_x$, 
where $H_x$ only acts on the variables at the sites $x$ and $x+1$.\footnote{$H_x = H_{\{ x,x+1 \} }$ in the notation introduced in \eqref{set dependence H A}.}
We also assume that $H_x$ is of the type $H_x = H^{(0)}_x + \coupling H^{(1)}_x$ with $\{ H^{(0)}_x, H^{(0)}_y \} = 0$ for all $x,y\in V$, 
where $\{ \cdot, \cdot \}$ denotes the Poisson bracket in classical mechanics, or the commutator in quantum mechanics. 
Given a site $a \in V$, we can define the energy current $G_{a,a+1}$ across the bond $(a,a+1)$ by 
\begin{align*}
\coupling G_{a,a+1} \; &= \; \Big\{ H , \sum_{x > a} H_x \Big\} \; = \; \{ H_a, H_{a+1}\} \\
\; &= \; \coupling \left( \{ H_a^{(0)}, H_{a+1}^{(1)}\} + \{ H_a^{(1)}, H_{a+1}^{(0)}\} \right) + \coupling^2 \{ H_a^{(1)}, H_{a+1}^{(1)}\} . 
\end{align*}
The Green-Kubo thermal conductivity is defined as the space-time variance of the current in equilibrium at temperature $T$:
\begin{equation}\label{Green Kubo}
\kappa (\coupling, T) \; = \; \lim_{t\to \infty} \lim_{|V|\to \infty} 
\frac{\coupling^2}{T^2}\left\langle
\left|
\frac{1}{\sqrt t} \int_0^t \frac{1}{\sqrt{|V|}} \sum_{a\in V} G_{a,a+1}(s) \, \dd s
\right|^2
\right\rangle_T,
\end{equation}
$\langle \cdot \rangle_T$ being the Gibbs state at temperature $T$, and $G_{a,a+1}(s)$ being the time evolution of $G_{a,a+1}$ at time $s$ under the action of the Hamiltonian.

We aim to understand the behavior of $\kappa (\coupling , T)$ in the integrable limit, being most notably the limit $\coupling \to 0$ for fixed $T\sim 1$, 
though it can also correspond to the limit $T\to 0$ or $T \to \infty$ at fixed $\coupling \sim 1$ for some of the examples considered below. 
Rigorous results on the conductivity of non-integrable Hamiltonian dynamics are most likely too hard to get at the present time. 
Nevertheless, in the examples below, we make the idea of asymptotic localization mathematically precise, 
by introducing some cut-off, possibly modeled by a energy-conserving noise, in the time integral in \eqref{Green Kubo}. 
We expect this cut-off to furnish a reasonable approximation to obtain an upper bound on $\kappa$, in both the scenarios where the dynamics is truly MBL or diffusive.

\paragraph{Classical disordered anharmonic chain.}
We consider a quenched disordered system as a first example. 
Despite the fact that translation invariance is broken, no true MBL phase is expected \cite{Dhar Lebowitz}\cite{Oganesyan Pal Huse}\cite{Basko}, 
and it is possible that rare mobile chaotic spots constitute the main delocalization mechanism.
Though we could allow for much more generality, let us take as an example the Hamiltonian 
\begin{equation}\label{anharmonic disordered hamiltonian}
H(q,p) \; = \; \sum_{x\in V} \left( p_x^2 + \omega_x^2 q_x^2 + \coupling (q_{x+1} - q_x)^2 + \coupling q_x^4 \right), 
\end{equation}
with the convention $q_{(|V|+1)/2} = q_{(|V|-1)/2}$, meaning that free boundary conditions are taken on both ends. 
We assume the frequencies $\omega_x$ to be i.i.d.\@, with smooth compactly supported density, bounded away from zero. 
Let $n\ge 1$ be some integer (we are interested in the case $n \gg 1$). 
As a cut-off in time, we perturb the Hamiltonian dynamics by an energy preserving noise that becomes relevant on time scales of order $\coupling^{-n}$, 
generating a very slow diffusive behavior. 
The generator of the full dynamics is given by 
\begin{eqnarray}
\mathcal L \; &=& \; L_H + \coupling^n S \qquad \text{with} \nonumber\\
 L_H \; &=& \; \{ H, \cdot \}
\quad \text{and} \quad S f (q,p) \; = \; \sum_{x\in V} \big( f(q, \dots, -p_x,\dots) - f(q,p) \big). \label{noisy}
\end{eqnarray}
Let $\tilde\kappa (\coupling,T,n)$ be the Green-Kubo conductivity defined for this noisy dynamics. 
In \cite{Huveneers}, we show
\begin{theorem}\label{Theorem: disordered}
For given $T> 0$, and for any integer $n\ge 1$, there exists a constant $\Const_{T,n}$ such that, for almost all realizations of the frequencies $\omega_x$, 
it holds that $\tilde\kappa (\coupling,T,n) \le \Const_{T,n} \coupling^n$. 
\end{theorem}
This result suggests that $\kappa (\coupling,T) = \mathcal O (\coupling^n)$ for any $n\ge 1$ as $\coupling \to 0$. 
We expect that the theorem remains true if $\coupling (q_{x+1} - q_x)^2$ is replaced by $(q_{x+1} - q_x)^2$ in \eqref{anharmonic disordered hamiltonian}, 
since all the eigenmodes of the chain are localized in the absence of nonlinearity (in $d=1$ or even in $d>1$ at strong enough disorder).
In that case, as the scaling $\kappa (\coupling, T) = \kappa (\coupling/r,r T)$ holds then for any $r>0$, 
we conclude that asymptotic localization at small coupling implies also asymptotic localization at low temperature. 

\paragraph{Classical rotors and DNLS chain.}
In \cite{1}, Theorem \ref{Theorem: disordered} is extended to the translation invariant rotors chain \eqref{rotor} with the same noise as in \eqref{noisy}. 
In this case, the scaling for the true conductivity is given by $\kappa (\coupling ,T) = \frac{1}{r}\kappa (r^2 \coupling, r^2 T)$ for any $r>0$, 
so that, this time, asymptotic localization at small coupling implies asymptotic localization at large temperature, in agreement with the observations of \cite{Giardina}. 

The Discrete Non-Linear Schr\"odinger (DNLS) chain, with Hamiltonian
\begin{equation}\label{DNLS}
H(\psi, \overline{\psi}) \; = \; \sum_{x\in V} \left( |\psi_x|^4 + \coupling |\psi_{x+1} - \psi_x|^2 \right), \qquad \psi_x \in \C
\end{equation}
with again $\psi_{(|V|+1)/2} = \psi_{(|V|- 1)/2}$, displays a thermal behavior similar to that of the rotors chain in the small coupling limit. 
Nevertheless, the introduction of an energy preserving noise is non-obvious. We adopt therefore a different, less physical, type of time cut-off.
For an integer $n\ge 1$, we define
\begin{equation*}
\hat\kappa (\coupling, T,n) \; = \; \lim_{|V|\to \infty} 
\frac{\coupling^2}{T^2}\left\langle
\left|
\frac{1}{\sqrt{\coupling^{-n}}} \int_0^{\coupling^{-n}} \frac{1}{\sqrt{|V|}} \sum_{a\in V} G_{a,a+1}(s) \, \dd s
\right|^2
\right\rangle_T, 
\end{equation*} 
and in \cite{1}, we prove 
\begin{theorem}
For given $T> 0$, 
and for the Hamiltonian dynamics generated by the Hamiltonian \eqref{DNLS}, it holds that $\lim_{\coupling \to 0} \coupling^{-(n-1)} \hat\kappa (\coupling, T,n) = 0$ for any integer $n\ge 1$. 
\end{theorem}
Again, this result suggests that $\kappa (\coupling,T) = \mathcal O (\coupling^n)$ for any $n\ge 1$ as $\coupling \to 0$. 
It is observed that asymptotic localization is not to be attributed to any statistical interplay with a second conserved quantity (total momentum for the rotors chain, 
total $\ell^2$-norm for the DNLS chain). The conservation of these quantities could be broken by an additional perturbation.  
We conjecture that this result stays valid in higher dimensions and for many more similar Hamiltonians, 
as for example $H(q,p) = \sum_x \{ p^2_x + q^4_x + \coupling (q_{x+1} - q_x)\}$. 
This has however not been shown at the present time.

\paragraph{Quantum chain analogous to the Bose-Hubbard chain.}
In \cite{2}, we show that a quantum chain, analogous to the Bose-Hubbard chain,
becomes asymptotically MBL in the limit $T \to \infty$.\footnote{
This limit could be compared to the limit $\NumLevels \to \infty$, while keeping the temperature infinite, in the set-up of Section \ref{section: quenched and thermal}.}
The Hamiltonian is given by 
\begin{equation*}\label{Pseudo Bose Hubbard}
H \; = \; \sum_{x\in V} \left( (a_x^* a_x)^q + \coupling (a_x^* a_{x+1} + a_x a_{x+1}^*) \right), \qquad q > 2, 
\end{equation*}
where $a_x$ and $a_x^*$ are respectively bosonic annihilation and creation operators (without cut-off). 
The need for the constraint $q>2$ is seen as follows.\footnote{
A regime of asymptotic MBL could still be expected for $q=2$ if both $T\to \infty$ and $\coupling \to 0$.
}
The typical energy per site is $N_x^q \sim T$ (with $N_x = a_x^* a_x$), 
so that the typical energy difference due to hopping is given by $(N_x + 1)^q - N_x^q \sim N_x^{q-1} \sim T^{1- 1/q}$. 
The coupling is of order $\coupling N_x \sim \coupling T^{1/q}$.
Therefore, the ratio between coupling and energy difference is given by $\coupling T^{2/q - 1}$. 
This ratio needs to go to zero as $T\to \infty$ for resonances to become sparse in this limit, imposing $q> 2$.

For any integer $n\ge 1$, let us define 
\begin{equation*}
\hat\kappa (\coupling, T,n) \; = \; \lim_{|V|\to \infty} 
\frac{\coupling^2}{T^2}\left\langle
\left|
\frac{1}{\sqrt{T^{n}}} \int_0^{T^{n}} \frac{1}{\sqrt{|V|}} \sum_{a\in V} G_{a,a+1}(s) \, \dd s
\right|^2
\right\rangle_T.
\end{equation*} 
The following theorem is shown in \cite{2}.
\begin{theorem}
There exists $\const < + \infty$ such that,
for the Hamiltonian dynamics generated by the Hamiltonian \eqref{Pseudo Bose Hubbard}, and for fixed $\coupling > 0$,
it holds that $\lim_{T \to \infty} T^{n - \const} \hat\kappa (\coupling, T,n) = 0$ for any integer $n\ge 1$. 
\end{theorem}
Let us notice that the conductivity depends separately on the temperature $T$ and the coupling strength $\coupling$. 
Our result does not imply asymptotic localization as $\coupling \to 0$ for a fixed temperature $T \sim 1$. 
Actually, especially for $d\ge 2$, the considerations of Section \ref{rare bubbles perturbative} 
lead us to think that the conductivity behaves as a power law in $\coupling$ in the limit $\coupling \to 0$.

\paragraph{Acknowledgements:}
We thank D.~Huse, J.~Imbrie, R.~Nandkishore, M.~M\"uller and M.~Schiulaz for discussions.

\section*{Appendix 1}
\addcontentsline{toc}{section}{Appendix 1}
We show a sequence of possible steps to connect $| \eta \rangle$ to $| \eta' \rangle$ via resonant transitions on figure \ref{figure: resonances tranlsation invariant Hamiltonian}. 
We start from $| \eta \rangle$ in the upper left corner, go from left to right and from top to bottom, and end up with $| \eta' \rangle$ in the lower right corner. 
Occupation numbers marked in red are the ones that will get swapped.

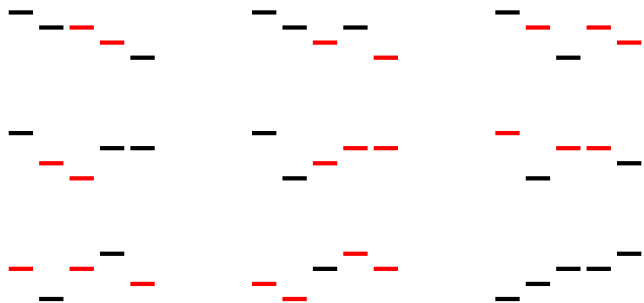
\begin{center}
\begin{tikzpicture}[scale=0.8]

\draw [ultra thick]  (0.05, 1) -- (0.45,1);
\draw [ultra thick]  (0.55, 0.75) -- (0.95,0.75);
\draw [ultra thick,red]  (1.05, 0.75) -- (1.45,0.75);
\draw [ultra thick,red]  (1.55, 0.5) -- (1.95,0.5);
\draw [ultra thick]  (2.05, 0.25) -- (2.45,0.25);

\begin{scope}[xshift=4cm]
\draw [ultra thick]  (0.05, 1) -- (0.45,1);
\draw [ultra thick]  (0.55, 0.75) -- (0.95,0.75);
\draw [ultra thick,red]  (1.05, 0.5) -- (1.45,0.5);
\draw [ultra thick]  (1.55, 0.75) -- (1.95,0.75);
\draw [ultra thick,red]  (2.05, 0.25) -- (2.45,0.25);
\end{scope}

\begin{scope}[xshift=8cm]
\draw [ultra thick]  (0.05, 1) -- (0.45,1);
\draw [ultra thick,red]  (0.55, 0.75) -- (0.95,0.75);
\draw [ultra thick]  (1.05, 0.25) -- (1.45,0.25);
\draw [ultra thick,red]  (1.55, 0.75) -- (1.95,0.75);
\draw [ultra thick,red]  (2.05, 0.5) -- (2.45,0.5);
\end{scope}

\begin{scope}[xshift=0cm,yshift=-2cm]
\draw [ultra thick]  (0.05, 1) -- (0.45,1);
\draw [ultra thick,red]  (0.55, 0.5) -- (0.95,0.5);
\draw [ultra thick,red]  (1.05, 0.25) -- (1.45,0.25);
\draw [ultra thick]  (1.55, 0.75) -- (1.95,0.75);
\draw [ultra thick]  (2.05, 0.75) -- (2.45,0.75);
\end{scope}

\begin{scope}[xshift=4cm,yshift=-2cm]
\draw [ultra thick]  (0.05, 1) -- (0.45,1);
\draw [ultra thick]  (0.55, 0.25) -- (0.95,0.25);
\draw [ultra thick,red]  (1.05, 0.5) -- (1.45,0.5);
\draw [ultra thick,red]  (1.55, 0.75) -- (1.95,0.75);
\draw [ultra thick,red]  (2.05, 0.75) -- (2.45,0.75);
\end{scope}

\begin{scope}[xshift=8cm,yshift=-2cm]
\draw [ultra thick,red]  (0.05, 1) -- (0.45,1);
\draw [ultra thick]  (0.55, 0.25) -- (0.95,0.25);
\draw [ultra thick,red]  (1.05, 0.75) -- (1.45,0.75);
\draw [ultra thick,red]  (1.55, 0.75) -- (1.95,0.75);
\draw [ultra thick]  (2.05, 0.5) -- (2.45,0.5);
\end{scope}

\begin{scope}[xshift=0cm,yshift=-4cm]
\draw [ultra thick,red]  (0.05, 0.75) -- (0.45,0.75);
\draw [ultra thick]  (0.55, 0.25) -- (0.95,0.25);
\draw [ultra thick,red]  (1.05, 0.75) -- (1.45,0.75);
\draw [ultra thick]  (1.55, 1) -- (1.95,1);
\draw [ultra thick,red]  (2.05, 0.5) -- (2.45,0.5);
\end{scope}

\begin{scope}[xshift=4cm,yshift=-4cm]
\draw [ultra thick,red]  (0.05, 0.5) -- (0.45,0.5);
\draw [ultra thick,red]  (0.55, 0.25) -- (0.95,0.25);
\draw [ultra thick]  (1.05, 0.75) -- (1.45,0.75);
\draw [ultra thick,red]  (1.55, 1) -- (1.95,1);
\draw [ultra thick,red]  (2.05, 0.75) -- (2.45,0.75);
\end{scope}

\begin{scope}[xshift=8cm,yshift=-4cm]
\draw [ultra thick]  (0.05, 0.25) -- (0.45,0.25);
\draw [ultra thick]  (0.55, 0.5) -- (0.95,0.5);
\draw [ultra thick]  (1.05, 0.75) -- (1.45,0.75);
\draw [ultra thick]  (1.55, 0.75) -- (1.95,0.75);
\draw [ultra thick]  (2.05, 1) -- (2.45,1);
\end{scope}

\end{tikzpicture}
\end{center}

\section*{Appendix 2}
\addcontentsline{toc}{section}{Appendix 2}
We prove (\ref{carre de lu}-\ref{stability carre de lu}).
Consider an initial classical state $| \eta \rangle$ and another classical state $| \eta' \rangle$ such that $\langle \eta' | H^{(0)} | \eta \rangle \ne 0$. 
There are thus points $x,y$ satisfying $|x-y|_1 = 1$ and such that $\eta'_x = \eta_x + 1$ and $\eta'_y = \eta_y - 1$, while $\eta'_z = \eta_z$ for all $z\ne x,y$. 
For concreteness, let us assume that $x=(x_1,x_2)$ and $y=(x_1,x_2+1)$ (other cases analogous), 
let us write $\eta_x=a$, $\eta_y=b$ and let us denote by $r,s, \dots, w$ the occupation numbers on the neighboring sites. 
The transition from $|\eta \rangle$ to $| \eta' \rangle$ is represented as
\begin{center}
\begin{tikzpicture}[scale=1]

\draw[fill] (0,2) circle(0.05); \draw(0,2) node[above]{$r$};
\draw[fill] (0,1) circle(0.05); \draw(0,1) node[above]{$b$};
\draw[fill] (0,0) circle(0.05); \draw(0,0) node[above]{$a$};
\draw[fill] (0,-1) circle(0.05); \draw(0,-1) node[above]{$u$};

\draw[fill] (-1,1) circle(0.05); \draw(-1,1) node[above]{$w$};
\draw[fill] (-1,0) circle(0.05); \draw(-1,0) node[above]{$v$};

\draw[fill] (1,1) circle(0.05); \draw(1,1) node[above]{$s$};
\draw[fill] (1,0) circle(0.05); \draw(1,0) node[above]{$t$};

\draw[->,>=latex] (2,0.5) -- (4,0.5);

\begin{scope}[xshift=6cm]
\draw[fill] (0,2) circle(0.05); \draw(0,2) node[above]{$r$};
\draw[fill] (0,1) circle(0.05); \draw(0,1) node[above]{$b-1$};
\draw[fill] (0,0) circle(0.05); \draw(0,0) node[above]{$a+1$};
\draw[fill] (0,-1) circle(0.05); \draw(0,-1) node[above]{$u$};

\draw[fill] (-1,1) circle(0.05); \draw(-1,1) node[above]{$w$};
\draw[fill] (-1,0) circle(0.05); \draw(-1,0) node[above]{$v$};

\draw[fill] (1,1) circle(0.05); \draw(1,1) node[above]{$s$};
\draw[fill] (1,0) circle(0.05); \draw(1,0) node[above]{$t$};
\end{scope}
\end{tikzpicture}
\end{center}
Because transitions must preserve the interaction energy according to \eqref{H 0 bubble model}, and because of the genericity condition \eqref{genericity condition}, 
the states $| \eta \rangle$ and $| \eta' \rangle$ must contain ``the same oriented nearest neighbors pairs with the same multiplicity", 
meaning that all the patterns $(w,b)$, $(b,s)$, $(v,a)$, $(a,t)$, $\binom{r}{b}$, $\binom{b}{a}$, $\binom{a}{u}$ must be found in $\eta'$. 
This imposes strong constraints on $| \eta \rangle$ for the transition to be possible. 
First we find that we need to have $b=a+1$, and then that $| \eta \rangle$ must actually be a configuration of one of the following two types
\begin{center}
\begin{tikzpicture}[scale=1]

\draw[fill] (0,2) circle(0.05); \draw(0,2) node[above]{$a+1$};
\draw[fill] (0,1) circle(0.05); \draw(0,1) node[above]{$a+1$};
\draw[fill] (0,0) circle(0.05); \draw(0,0) node[above]{$a$};
\draw[fill] (0,-1) circle(0.05); \draw(0,-1) node[above]{$a+1$};

\draw[fill] (-1,1) circle(0.05); \draw(-1,1) node[above]{$v$};
\draw[fill] (-1,0) circle(0.05); \draw(-1,0) node[above]{$v$};

\draw[fill] (1,1) circle(0.05); \draw(1,1) node[above]{$s$};
\draw[fill] (1,0) circle(0.05); \draw(1,0) node[above]{$s$};

\draw(3,0.5) node{or} ;

\begin{scope}[xshift=6cm]
\draw[fill] (0,2) circle(0.05); \draw(0,2) node[above]{$a$};
\draw[fill] (0,1) circle(0.05); \draw(0,1) node[above]{$a+1$};
\draw[fill] (0,0) circle(0.05); \draw(0,0) node[above]{$a$};
\draw[fill] (0,-1) circle(0.05); \draw(0,-1) node[above]{$a$};

\draw[fill] (-1,1) circle(0.05); \draw(-1,1) node[above]{$v$};
\draw[fill] (-1,0) circle(0.05); \draw(-1,0) node[above]{$v$};

\draw[fill] (1,1) circle(0.05); \draw(1,1) node[above]{$s$};
\draw[fill] (1,0) circle(0.05); \draw(1,0) node[above]{$s$};
\end{scope}
\end{tikzpicture}
\end{center}
for some $s$ and $v$. 
Now, for $| \eta \rangle$ satisfying \eqref{carre de lu}, we find a configuration depicted as 
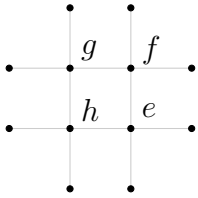
\begin{center}
\begin{tikzpicture}[scale=0.8]

\draw[color=gray!40] (-1,0) -- (2,0);
\draw[color=gray!40] (-1,1) -- (2,1);

\draw[color=gray!40] (0,-1) -- (0,2);
\draw[color=gray!40] (1,-1) -- (1,2);

\draw[fill] (-1,1) circle(0.05); 
\draw[fill] (-1,0) circle(0.05); 

\draw[fill] (0,2) circle(0.05); 
\draw[fill] (0,1) circle(0.05); \draw(0,1.3) node[right]{$g$};
\draw[fill] (0,0) circle(0.05); \draw(0,0.3) node[right]{$h$};
\draw[fill] (0,-1) circle(0.05); 

\draw[fill] (1,2) circle(0.05); 
\draw[fill] (1,1) circle(0.05); \draw(1,1.3) node[right]{$f$};
\draw[fill] (1,0) circle(0.05); \draw(1,0.3) node[right]{$e$};
\draw[fill] (1,-1) circle(0.05); 

\draw[fill] (2,1) circle(0.05);
\draw[fill] (2,0) circle(0.05);

\draw (4,0.5) node[right]{$|g-f| \ge 2$, $|f-e|\ge 2$, $|e-h|\ge 2$, $|h-g|\ge 2$.};
\end{tikzpicture}
\end{center}
In order for $| \eta \rangle$ to be connected by $H^{(0)}$ to a state $|\eta' \rangle$ where at least one of the values $e,f,g$ or $h$ has changed, 
a particle must hop along one of the twelve bonds appearing on the last figure. By the above, this is not possible. This shows \eqref{stability carre de lu}.


\end{document}